\documentclass[bibyear]{aa}
\usepackage{natbib}
\usepackage{multicol}
\usepackage{color}
\usepackage{verbatim}
\usepackage{graphicx}
\usepackage{txfonts}
\usepackage{supertabular,booktabs}
\usepackage{lscape}
\newcommand{\dgr}{$^{\circ}$}
\newcommand{\ecl}{$\beta$}

\begin{document}

\title{Sun-related variability in the light curves of compact radio sources}
\subtitle{A new view on extreme scattering events}

   \author{N. Marchili\inst{1}
          \and
         G. Witzel\inst{2}
         \and
         M. F. Aller\inst{3}
          }

   \institute{Istituto di Radioastronomia - INAF, Via Piero Gobetti 101, 40129, Bologna (Italy)\\
              \email{nicola.marchili@inaf.it}
         \and
             Max-Planck Institut fuer Radioastronomie, Auf del H\"ugel
             69, 53121, Bonn, Deutschland\\
             \email{gwitzel@mpifr-bonn.mpg.de}
         \and	
             Department of Astronomy, University of Michigan, 
             323 West Hall, Ann Arbor, MI 48109-1107, USA\\
             \email{mfa@umich.edu}
             }

   \date{Received ; accepted }

% \abstract{}{}{}{}{} 
% 5 {} token are mandatory
 
\abstract
   {Compact radio sources can show remarkable flux density variations at GHz frequencies on a wide range of timescales. The origin of the variability is a mix of source-intrinsic mechanisms and propagation effects, the latter being generally identified with scattering from the interstellar medium.
   Some of the most extreme episodes of variability, however, show characteristics that are not consistent with any of the explanations commonly proposed.}
   {An in-depth analysis of variability at radio frequencies has been carried out on light curves from the impressive database of the US Navy's extragalactic source monitoring program at the Green Bank Interferometer (GBI) --- a long-term project mainly aimed at the investigation of extreme scattering events --- complemented by UMRAO light curves for selected sources. The purpose of the present work is to identify events of flux density variations that appear to correlate with the position of the Sun.}
   {The 2 GHz and 8 GHz light curves observed in the framework of the GBI monitoring program have been inspected in a search for one-year periodic patterns in the data. Variations on timescales below one year have been isolated through a de-trending algorithm and analysed, looking for possible correlations with the Sun's position relative to the sources.}
   {Objects at ecliptic latitude below $\sim20^\circ$ show one-year periodic drops in flux densities, centred close to the time of minimum solar elongation; both interplanetary scintillation and instrumental effects may contribute to these events. However, in some cases the drops extend to much larger angular distances, affecting sources at high ecliptic latitudes, and causing variability on timescales of months. Three different kinds of such events have been identified in the data. Their exact nature is not yet known.} 
   {In the present study we show that the variability of compact radio sources is heavily influenced by effects that correlate with solar angular distance; this unexpected contribution significantly alters the sources' variability characteristics estimated at GHz frequencies. In particular, we found that many extreme scattering events previously identified in the GBI monitoring program are actually the consequence of Sun-related effects; others occur simultaneously in several objects, which excludes interstellar scattering as their possible cause. These discoveries have a severe impact on our understanding of extreme scattering events. 
   Furthermore, Sun-related variability, given its amplitude and timescale, can significantly alter results of variability studies, which are very powerful tools for the investigation of active galactic nuclei. Without a thorough comprehension of the mechanisms that cause these variations, the estimation of some essential information about the emitting regions, such as their size and all the derived quantities, might be seriously compromised.}
   

   \maketitle
%
%________________________________________________________________

\section{Introduction}

Variability studies provide us with a powerful tool to investigate the nature of the emission from extragalactic compact radio sources. For these objects, variability can appear over a wide range of timescales, going from hours (intrahour/intraday variability, hereafter IDV; see \citealt{1986MitAG..65..239W}, \citealt{1987AJ.....94.1493H}) to many years (see e.g. \citealt{1997ARA&A..35..445U}).

It is generally assumed that the origin of the observed flux density variations in the radio is either intrinsic to the sources or caused by interstellar scintillation (ISS; see, e.g., \citealt{1995ARA&A..33..163W}; \citealt{2006ApJS..165..439R}), a propagation effect due to the presence of a screen of interstellar medium (ISM) along the line of sight between the source and the observer. The importance of scattering in the ISM on centimetre band data was demonstrated, among others, by \cite{2015MNRAS.452.4274P} and \cite{2022MNRAS.515.1736K}. In \cite{2011A&A...530A.129M} it was shown that, on IDV timescales, a further contribution to the variability is provided by the Sun, either through local propagation effects (i.e. interplanetary scintillation, IPS), or, indirectly, through instrumental effects rising when a source's angular distance to the Sun, that is its solar elongation, is low. The analysis of 5\,GHz light curves from the Urumqi Observatory and the 4.9\,GHz ones from the MASIV survey (\citealt{2003AJ....126.1699L}) at the Very Large Array (VLA) revealed a significant increase of the intraday variability (up to a factor 2) of the analysed sources as their solar elongation decreases. The aim of the present study is to assess how the Sun can affect the variability curves of compact radio sources on longer timescales, from a few days up to one year. This goal was pursued by searching for variability features that cannot be explained in terms of source-intrinsic processes or through ISS, such as solar-elongation-dependent flux density variations or {\it correlated} variability among many objects. Given the variety of manifestations of the effects we found, the results of our analysis have been split in two publications; the present one (Paper I) deals with sharp flux density variations similar to extreme scattering events (ESEs, \citealt{1987Natur.326..675F}). The second publication (Witzel et al., in prep.; from now on, Paper II) focuses on smooth one-year periodic variations in radio light curves. At this stage of the work, it is not known whether and how the two phenomena are related.
To reach our goal, we inspected one of the best available databases for the study of compact objects' variability at GHz frequencies, namely the US Navy's extragalactic source monitoring program  (henceforth, NESMP; see \citealt{1987ApJS...65..319F}, \citealt{1991ApJS...77..379W}, and \citealt{2001ApJS..136..265L}).

The NESMP was a project running at the Green Bank Interferometer (GBI) between 1979 and 1996, whose primary aim was to search for ESEs. These are a class of dramatic variations in the flux density of compact radio sources, characterised by a substantial decrease bracketed by less pronounced increases. Extreme scsttering events have a typical timescale between a few weeks and several months; their origin is generally attributed to the scattering of radio waves in the interstellar medium. These events are rare: \cite{1994ApJ...430..581F} estimated that, taking into account a cumulative observing time of about 600 source-years for the 330 objects they monitored, the timespan covered by unusual variability associated to ESEs amounts to 4.8 y, corresponding to about 0.8\% of the total time.

A thorough analysis of Sun-related variability (from now on, SRV) is crucial on several levels. Given the importance of variability studies to constrain basic properties of emitting sources (e.g. emitting regions' sizes, and, from these, brightness temperatures and Doppler factors; \citealt{2009A&A...494..527H}, \citealt{2017MNRAS.466.4625L}), and to understand the origin and the evolution of flares (see, e.g., \citealt{1992vob..conf...85M}, \citealt{1997ARA&A..35..445U}, \citealt{2016Galax...4...37M}), it is necessary to identify and quantify all non-intrinsic sources of variability in the light curves; the contribution of SRV to blazar variability at GHz frequencies seems to be quite significant. The nature of SRV is a puzzling and fascinating topic: the evidence we collected rules out an instrumental origin at least for some of the SRV manifestations. Propagation effects seem to be the most obvious candidate to explain them, but, in the light of the known properties of IPS or of the Earth's upper atmospheric layers, these effects would be expected to be generally negligible; this implies that our picture of local propagation effects must be missing some important pieces. Last but not least, SRV has a strong impact on our understanding of ESEs, given the fact that their signature in the light curves appears to be the same.

After a general description of the database and the procedures used for the data analysis (Sect. \ref{sec:data}), we will present the four different kinds of SRV manifestations we identified in the data (Sec \ref{sec:SRV All}) and we will illustrate our classification of the sources (Sect. \ref{sec:cls}). Some hypotheses about the origin of the variability will be discussed in Sect. \ref{sec:discussion}, while Sect. \ref{sec:eses} will be dedicated to a revision of the identified ESEs in \cite{2001ApJS..136..265L}, in the light of the new evidence we collected. The main conclusions of the present work will be presented in Sect. \ref{sec:concl}.

\section{Data and procedures}
\label{sec:data}

The NESMP observations were performed at two radio frequencies, of approximately 2.5 and 8.2\,GHz. Flux density measurements were collected for 149 objects, with a typical cadence of one observation every two days. Information about the GBI can be found in \cite{1969AJ.....74.1206H} and \cite{1973IEEEP..61.1335C}. The procedures for the data calibration and the results of the data analysis have been thoroughly discussed in \cite{1987ApJS...65..319F},  \cite{1991ApJS...77..379W}, and \cite{2001ApJS..136..265L}. Here below a short summary of the main aspects regarding data acquisition and calibration is presented.

Observations were carried out on a 2.4 km baseline, initially at frequencies of 2.7\,GHz (S band) and 8.1 GHz (X band). In September 1989 the installation of cryogenic receivers brought a change of the observed frequencies, which became 2.25\,GHz and 8.3\,GHz respectively. Four sources were used for the calibration of raw data; 1328+307 was used to fix the individual flux densities of 0237-233, 1245-197, 1328+254; 
the flux density measurements of these three sources were combined to form a hybrid calibrator for all the other targets. A deeper look into this procedure will be given in Appendix \ref{sec:App1}. According to \cite{1987ApJS...65..319F}, the uncertainties of GBI's flux density measurements before the installation of the cryogenic receivers could be expressed as

\begin{equation}
{\sigma_s}^2 = (0.037\ \mathrm{Jy})^2 + (0.014\ S_\mathrm{s})^2
\end{equation}
\begin{equation}
{\sigma_x}^2 = (0.057\ \mathrm{Jy})^2 + (0.049\ S_\mathrm{x})^2
\end{equation}

\noindent while after completion of the operation (August 1989), they could be expressed (see \citealt{2001ApJS..136..265L}) as

\begin{equation}
{\sigma_s}^2 = (0.0037\ \mathrm{Jy})^2 + (0.015 S_\mathrm{s})^2\\
\end{equation}
\begin{equation}
{\sigma_x}^2 = (0.0057\ \mathrm{Jy})^2 + (0.05 S_\mathrm{x})^2
\end{equation}

\noindent where $S_\mathrm{s}$ and $S_\mathrm{x}$ indicate the measured flux densities in S and X band respectively. A further contribution to the uncertainties, of the order of 10 and 20\% (for S and X band respectively), could be explained by atmospheric and hardware effects.

Some basic information about the analysed sources are reported in Table \ref{tab:allInfo}, such as name (Col. 1), ecliptic latitude (Col. 2), time of minimum solar elongation (Col. 3), average flux density and standard deviations at 2 and 8 GHz (Col. 4-7). The X-ray binary system 1909+048 has been excluded from the present analysis, as its source-intrinsic variability component is so strong to be absolutely dominating over any other kind of variability.

For a deeper study of SRV episodes in the blazars OJ\,287 and 0528+134, we made use of 8\,GHz and 14.5\,GHz data from the University of Michigan Radio Astronomical Observatory (UMRAO; for more details see, for example, \citealt{1985ApJS...59..513A}, \citealt{1999ApJ...512..601A}), which allowed for both an extension of the analysis to higher frequencies and an essential consistency check between the 8\,GHz data from the two facilities.

\subsection{Algorithms for the data analysis}
\label{sec:algorithms}

Two main procedures have been used for the analysis of the data: a de-trending procedure to remove the long-term variability and highlight the fast variations; a data-stacking procedure for identifying yearly patterns in the data. 

The de-trending procedure (see \citealt{2002A&A...390..407V}) consists of several steps: firstly, the flux densities of a light curve are averaged on time intervals of selected length; the resulting average fluxes are fitted by a spline curve, which is then interpolated with the same sampling of the original light curve; the re-sampled spline curve (i.e. the long-term trend) is subtracted from the original flux densities, leaving only the variations on timescales smaller than the time intervals used for the initial data averaging. 

The data-stacking procedure makes use of the de-trended data: these are divided into intervals of one year, which then are stacked on top of each other; the resulting variability curve as a function of day-of-the-year is finally averaged on bins of four days\footnote{The size of the bins has been chosen as a compromise between temporal resolution and number of datapoints per bin.
%the need of having a reasonable number of datapoints per bin, and the aim of a high temporal resolution
}, providing us with a mean yearly pattern of the fast variability.

\begin{figure}
   \centering
   \includegraphics[width=0.85\columnwidth]{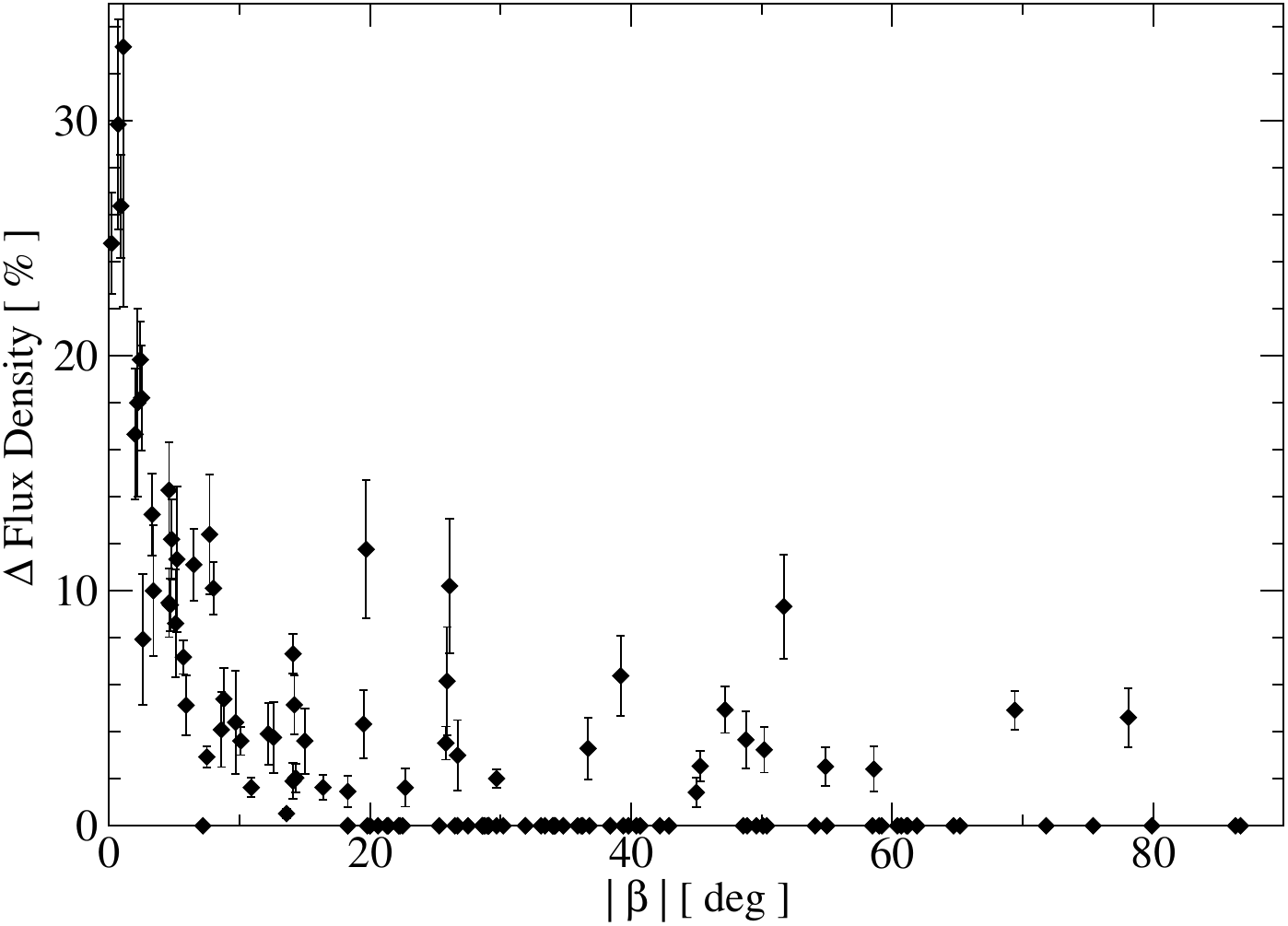}
      \caption{
	The average amplitude (expressed in percentage of the mean flux density) of the systematic dips at 2 GHz is plotted, for each source, versus \ecl. Particularly remarkable is the existence of sources showing SRV at latitudes higher than 20\dgr.}
      \label{fig:dips}
\end{figure}

\section{SRV phenomenology}
\label{sec:SRV All}

The light curve of some NESMP sources show one-year periodic dips at the time of minimum solar elongation, which were explained by \cite{1987ApJS...65..319F} as due to the effect of the Sun passing through a sidelobe of the primary antenna beam. 

In order to identify and characterise the periodic variations in the data, we applied the data-stacking procedure (with a de-trending interpolation timescale of ten months) to all sources in the NESMP catalogue. Along with the fast variability (with timescale from days to a few weeks) at very low solar elongation mentioned above, this analysis revealed three more kinds of systematics: slow (timescale of months) one-year periodic variability, correlated with solar elongation, mainly affecting sources at low ($<20$\dgr) ecliptic latitude; slow six-month periodic variability correlated with solar elongation; six-month periodic variability apparently not correlated with solar elongation. In all four cases (which will be separately discussed in the following), the flux density variations have shapes consistent with an ESE.

For sources showing flux density variations that resemble an ESE, centred at the time of minimum solar elongation, we fitted a gaussian profile --- centred at minimum solar elongation too --- to the stacked 2 GHz data. For each source, the amplitude and uncertainty of the fit have been used to quantify the dips' characteristics: these values are reported in Table \ref{tab:allInfo}, Col. 8. The dips' amplitudes are plotted in Fig. \ref{fig:dips}, divided by the average flux of the sources for a proper comparison. A clear dependence on ecliptic latitude, \ecl, can be seen up to \ecl$\sim20$\dgr; at higher latitudes, instead, the correlation disappears, which is a clear indication of a different mechanism producing the variability. 

It is important to stress that the vast majority of dips identified in the annual pattern of the data through the data-stacking procedure occurs either at minimum or at maximum solar elongation, which shows that the appearance of systematic dips in the light curves correlates with the position of the Sun with respect to the sources.

\begin{figure}
   \centering
   \includegraphics[width=0.85\columnwidth]{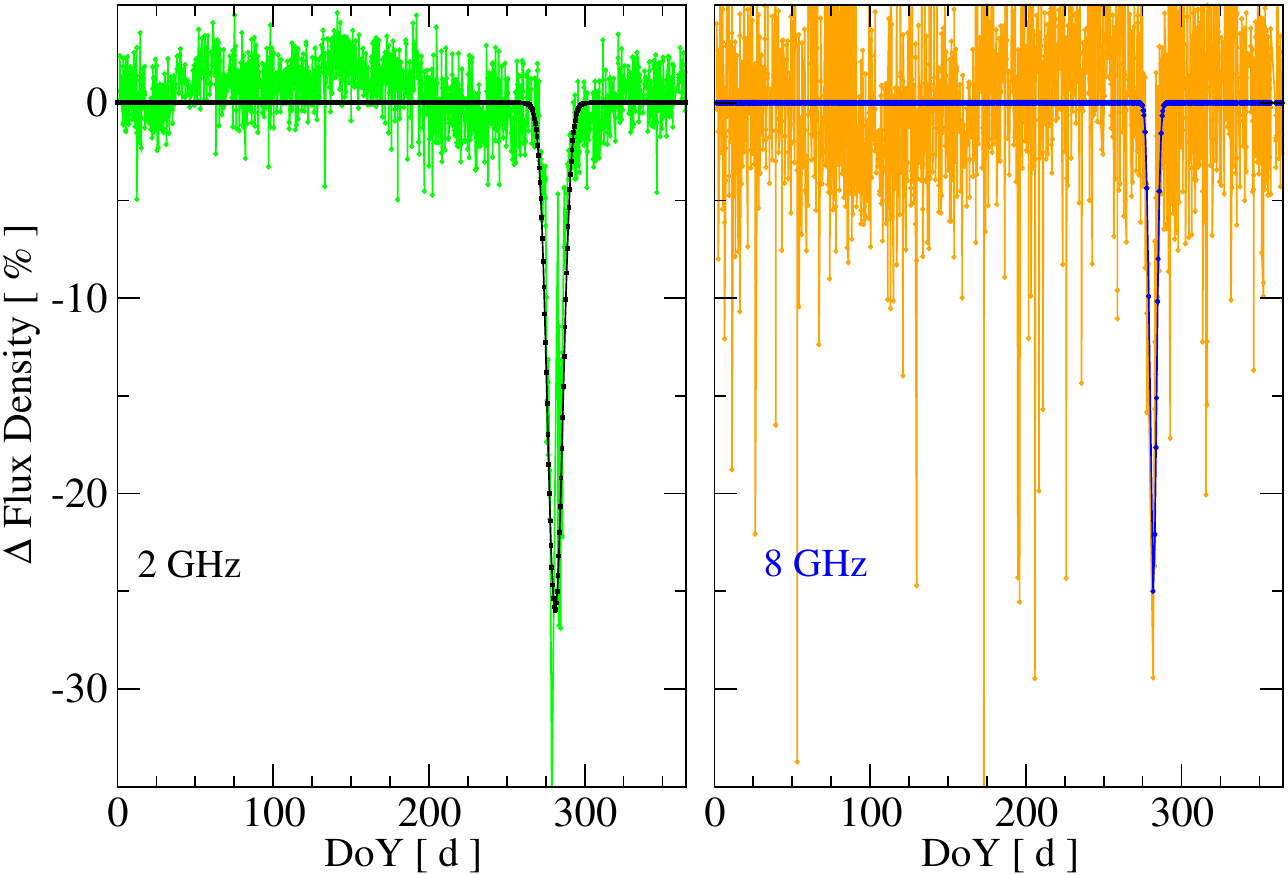}
      \caption{
	Seasonal flux density variations of 1253-055 at 2 and 8 GHz, calculated by de-trending the light curves on bins of ten months. The de-trended data are plotted versus day of the year (DoY); black and blue squares show the best fitting gaussian models. At 2 GHz, the effect of the Sun starts around Doy 265 and ends around DoY 298, indicating an influence spreading to solar elongations up to $\sim16^\circ$. For the 8 GHz data the influence is limited to  $\sim8^\circ$.}
      \label{fig:doy1}
\end{figure}

\subsection{Fast SRV at low ecliptic latitude (type I SRV)}
\label{sec:typeI}

The mean yearly pattern of 1253-055 shown in Fig. \ref{fig:doy1} nicely illustrates the typical effect of the Sun on sources at low ecliptic latitude at different frequencies. Systematic dropouts at 2 and at 8 GHz are comparable in amplitude, with flux density variations up to $\sim$30\%, but not in width: at 2 GHz data are affected up to a solar elongation of 15-20\dgr, while at 8 GHz the solar influence does not go further than 8\dgr.
The characteristics of this kind of SRV (which henceforth, for simplicity, will be denoted as type I SRV) are compatible with the explanation provided by \cite{1987ApJS...65..319F} in terms of an instrumental effect, due to the Sun passing through a sidelobe of the primary antenna beam. A further contribution to the variability may also come from IPS, whose effect is expected to become important at very low solar elongation.

\subsection{Annual SRV up to high solar elongation (type IIa SRV)}
\label{sec:OJ}

The annual pattern of some NESMP sources shows a systematic flux density decrease, correlated with solar elongation, extending over a very large interval of time. An example of this kind of variability is provided by the light curves of OJ~287. 

OJ~287 is among the best studied blazars; its fame grew considerably with the discovery of a long-term ($\sim12$\,y) periodic pattern in the optical band light curve (\citealt{1988ApJ...325..628S}). It has an ecliptic latitude of 2.6\dgr, which makes it a natural candidate to show type I SRV. We analysed the OJ~287 data from both the NESMP (2 and 8\,GHz) and the UMRAO (8 and 14.5\,GHz) monitoring campaigns, to check the consistency among flux density measurements with different telescopes, and to extend the investigation of slow SRV to higher frequencies.

\begin{figure}
   \centering
   \includegraphics[width=0.85\columnwidth]{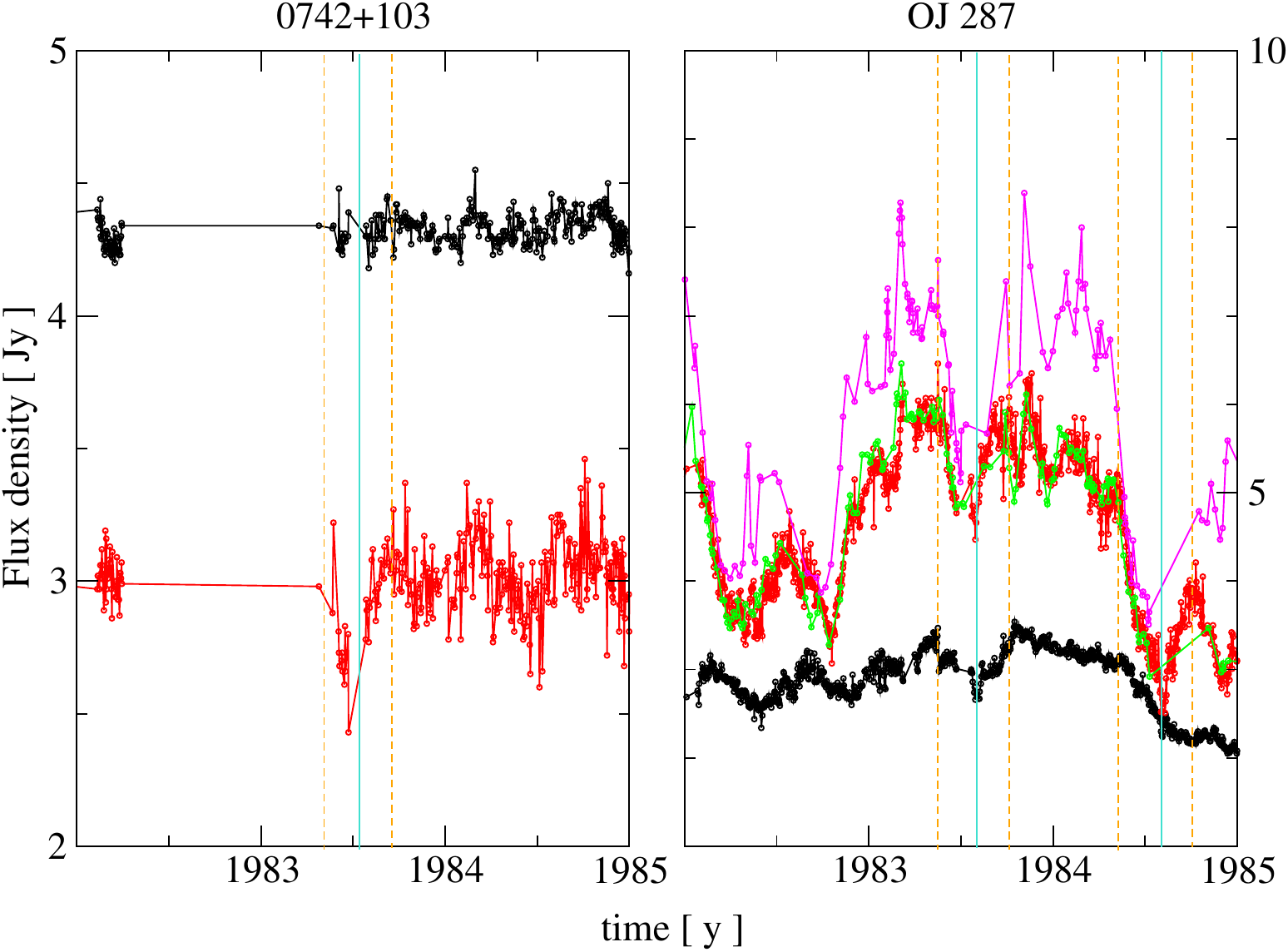}
      \caption{The 1983 type IIa SRV event in 0742+103 and OJ\,287 (left and right panel, respectively). The event is stronger at 8 (red dots) than at 2 GHz (black dots), and even stronger at 14.5 GHz (magenta dots, from the UMRAO archive). Note the excellent agreement between GBI and UMRAO 8 GHz data (green dots). 
      Turquoise lines mark the times of minimum solar elongation for the two sources, while orange lines indicate the beginning and end of the events.}
      \label{fig:0742-OJ}
\end{figure}

The data-stacking procedure reveals the existence of type I SRV at 2 GHz, as expected. Surprisingly, at 8 and 14.5 GHz the dip is much larger than at 2 GHz, as it extends from about $0.45$ to $0.70-0.75$\,y, which correspond to a solar elongation range up to $50-55$\dgr. The dips do not recur systematically every year, but they occur frequently: an example of two consecutive dips, in 1983 and 1984, is shown in Fig. \ref{fig:0742-OJ}, right panel. The two dips, which are almost identical in amplitude and duration, are stronger at 8 and 14.5 GHz than at 2 GHz. 

Particularly important is the simultaneity of the flux density variations in the different bands. Generally, the variability of OJ~287 shows spectral evolution: a correlation analysis of the light curves through locally normalised discrete correlation function (henceforth NDCF, see \citealt{1992ApJ...384..453L, 1988ApJ...333..646E}) reveals that the 14 GHz variations occur, on average, about 0.08 y earlier than at 8 GHz, and about 0.25 y earlier than at 2 GHz. At minimum solar elongation, however, the time delays between these frequencies drops to zero. Also remarkable, in Fig. \ref{fig:0742-OJ}, is the occurrence, in 1983, of a flux density drop in 0742+103 (\ecl=-10.9\dgr), similar in amplitude and duration to the one in OJ~287, and similarly stronger at high rather than at low frequencies (see Fig. \ref{fig:0742-OJ}, left panel). Given the small angular distance between the sources, it is very likely that these dips have the same origin. A weaker dip, partially hidden by the noise, is visible in the 0742+103 light curve in 1984 too, at 8 GHz.

More sources show evidence of the same broad systematic dips at minimum solar elongation found in OJ~287 (which henceforth we will address as type IIa SRV); they are indicated in Table \ref{tab:allInfo}, Col. 9 and 10, for the 2 and the 8 GHz light curves separately. Several of them are very close to OJ~287 in right ascension and declination. Type IIa SRV at 2 GHz is often associated with the same kind of variability at 8 GHz, although exceptions are not rare: in 2059+034, for example, two strong ESE-like variability episodes at minimum solar elongation can be found at 2 GHz (at 1990.13 and 1992.10), but not at 8 GHz.

\subsection{Semi-annual SRV up to high solar elongation (type IIb SRV)}
\label{sec:0537}

\begin{figure}
   \centering
   \includegraphics[width=0.85\columnwidth]{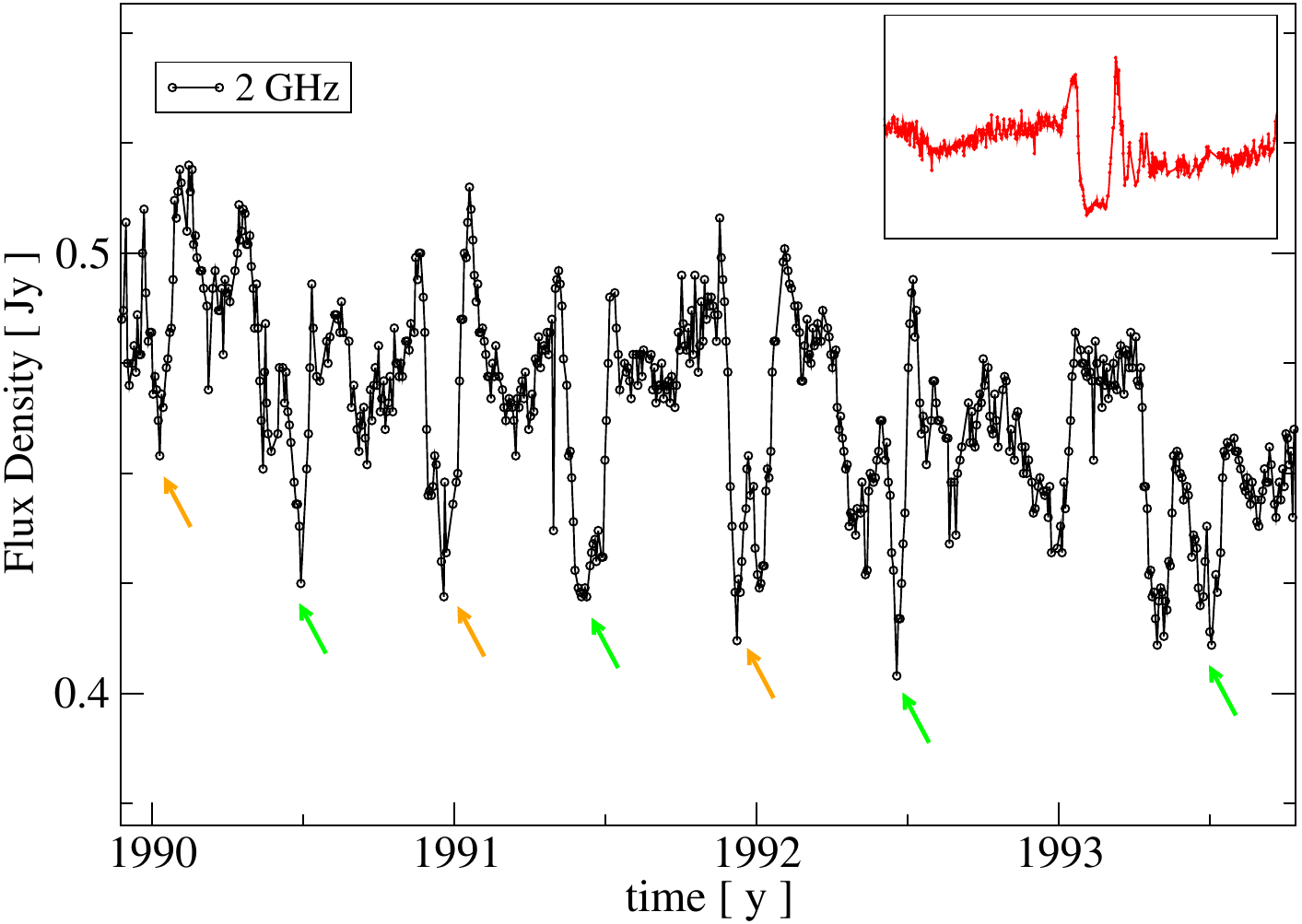}
      \caption{
      The 2 GHz light curve of 0537-158. Between 1990 and 1993 the variability of the source appears as a sequence of six-month separated ESEs, centred at the time of minimum and maximum solar elongation of the source (indicated, respectively, with green and orange arrows). In the upper right box, as a comparison, a zoom-in on the 2 GHz 1981 event of 0954+658, which can be considered as the archetype of ESEs.
}
      \label{fig:0537}
\end{figure}

A rather peculiar manifestation of solar-elongation-dependent variability is the existence of six-month periodic dips, similar to ESEs, in the light curves of some NESMP sources (indicated in Table \ref{tab:allInfo}, Col. 9 and 10); we will refer to it as type IIb SRV. The most striking example is given by the 2 GHz light curve of 0537-158 (\ecl = -39\dgr), which, between 1990 and 1994, appears as a regular sequence of ESEs with six-month cadence (see Fig. \ref{fig:0537}).

The average duration of the events is of the order of months. Type IIb SRV generally affects both 2 and 8 GHz light curves, although, similarly to the case of type IIa SRV, the way the two frequencies are affected varies from source to source: 1830+285, for example, doesn't show any variability resembling an ESE at 8 GHz. Also, the effect is not equally strong at minimum and at maximum solar elongation: generally the features at minimum are stronger than at maximum solar elongation, but in 0528+134 and 0954+658 the opposite occurs. 

\begin{figure}
   \centering
   \includegraphics[width=0.85\columnwidth]{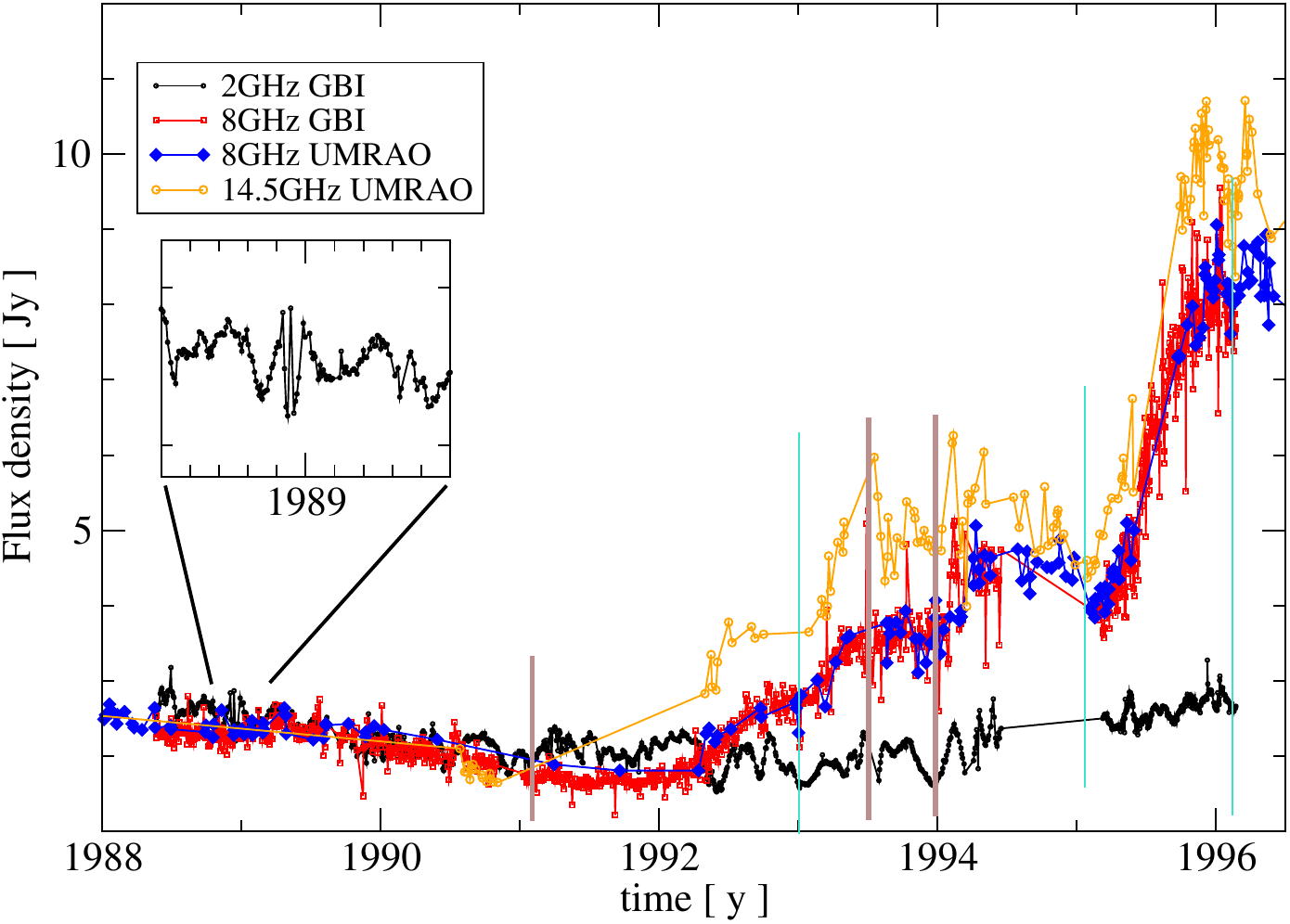}
      \caption{GBI and UMRAO light curves of 0528+134. Brown lines indicate the ESEs identified by \cite{2001ApJS..136..265L}, while turquoise ones show three more simultaneous multi-frequency dips recognisable in the data. In the small box the 1988.9 event, a rare example of complex ESE shape observable at 2 GHz.}
      \label{fig:0528}
\end{figure}

The case of 0528+134 deserves particular attention. In \cite{2001ApJS..136..265L}, the source is reported as showing three ESEs, in 1991.0, 1993.5, and 1993.9, which makes it the most affected object by ESEs in the whole monitoring program. All the reported events are consistent with the semi-annual cadence of type IIb SRV (the source reaches minimum solar elongation at 0.45 y). A comparison between the data collected within the NESMP with the ones from the UMRAO database (see Fig. \ref{fig:0528}) shows that the flux density measurements from the two facilities are in excellent agreement. 
Also in this case, it is important to stress the achromaticity of the flux density variations: in general, the variability of 0528+134 shows an overall trend of spectral evolution, with a delay, calculated via NDCF, of $\sim0.17$\,y between 14.5 and 8\,GHz data, and much larger between 8 and 2\,GHz. Most of the variability observed on timescales of one year or shorter, however, is achromatic (the 14.5-8\,GHz cross-correlation time delay of de-trended data is an impressive $0.00\pm0.05$ y). Along with the three ESEs identified by \cite{2001ApJS..136..265L}, indicated with brown lines, three more achromatic dips can be seen in the data (turquoise lines), namely in 1993.0, 1995.1, and 1996.1. The last two are particularly strong at 14.5\,GHz (orange dots). Another event, visible only at 2\,GHz, occurs in 1988.9 (see the small box within the figure); this is a rare example of a complex ESE shape detected at this frequency. The semi-annual nature of the variability, correlated with solar elongation, is very clear, which confirms its attribution to SRV. The agreement between NESMP and UMRAO data, and the importance of the effect at 14.5\,GHz, confirm the non-instrumental origin of type IIb SRV and its achromaticity within the range of investigated wavelengths.

\subsection{Time-dependent semi-annual variability (TDV)}
\label{sec:time}

So far, all the systematic variability we discussed showed a strict correlation to solar elongation, as the flux density minima occurred either at minimum or at maximum solar elongation. The NESMP data show a further interesting kind of ESE-like variability that is not directly ascribable to the relative position of the Sun to a source; it seems instead to be related to a specific time of the year. This kind of variability will be henceforth addressed as Time-Dependent Variability (TDV).

\begin{figure}
   \centering
   \includegraphics[width=0.85\columnwidth]{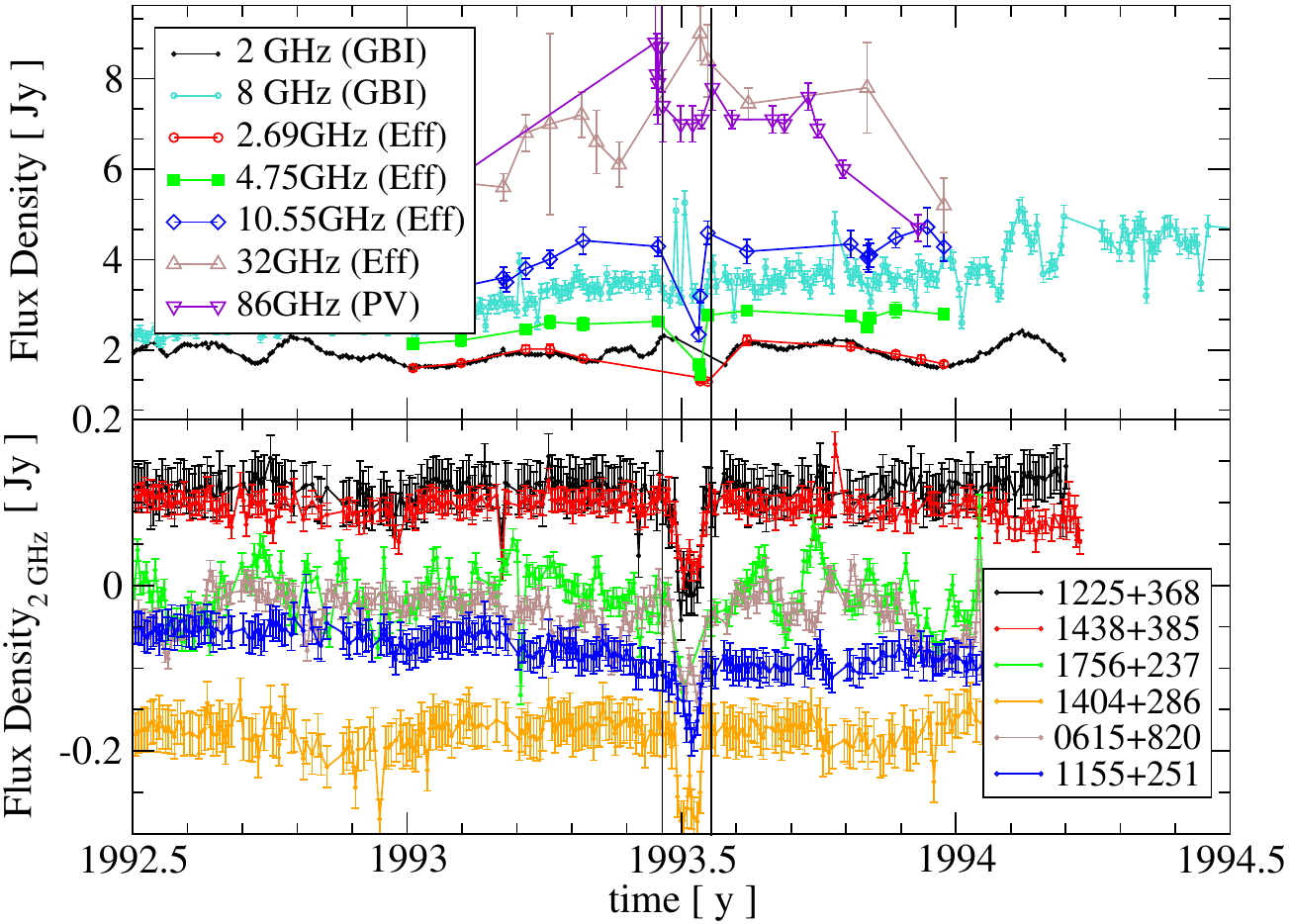}
      \caption{Upper panel: the 0528+134 ESE reported by \cite{1996A&AS..120C.529P} using data from the 100m-Effelsberg telescope and the 30m Pico Veleta one. Bottom panel: some examples of the ESE-like feature in the GBI light curves at the very same time. The light curves are de-trended and shifted in flux for an easier comparison.}
      \label{fig:July93}
\end{figure}

In Tables 5, 6, and 7 of \cite{2001ApJS..136..265L}, which present lists of possible ESEs from the GBI monitoring program, eight sources are reported as showing identified or potential ESEs \emph{approximately at the same time}, in 1993.5 (i.e. July 1993). A thorough check of the complete NESMP database reveals that a flux density drop compatible with the shape of an ESE is visible in many other sources. The light curves of some of them are plotted in the bottom panel of Fig. \ref{fig:July93}, after applying the de-trending procedure with time intervals of 50 days. 

This event is both extraordinary and puzzling for several reasons. Among the affected sources there is 0528+134, which has already been discussed as affected by type IIb SRV, with six-month periodic ESE-like variability close to the time of minimum and maximum solar elongation, occurring at 0.45 and 0.95\,y. The 1993.5 event occurs close to the minimum in solar elongation, therefore it is consistent with the solar-elongation-related semi-annual trail of events. For many other sources (e.g., 1225+368, 1404+286, and 1438+385), however, the time of the dip does not come close either to a minimum or to a maximum of solar elongation, which demonstrates that the 1993.5 event is truly time-dependent, rather than solar elongation-dependent. Furthermore, the 1993.5 ESE of 0528+134 has been detected by \cite{1995A&A...303..383P} in data from the 100-m Effelsberg telescope and the 30-m Pico Veleta telescope. This poses some questions: correlated variations simultaneously detected in many objects could point to an instrumental or a calibration problem, but how to explain the excellent consistency with the data from different telescopes? Also the interpretation of  the variability of 0528+134 is problematic: is it solar-elongation dependent or time-dependent? This point will be addressed in Sect. \ref{sec:discussion}, where we will discuss a possible link between the different manifestations of SRV and TDV. The hypothesis that TDV is due to a calibration problem is assessed in Appendix \ref{sec:App1}.

\begin{figure}
   \centering
   \includegraphics[width=0.85\columnwidth]{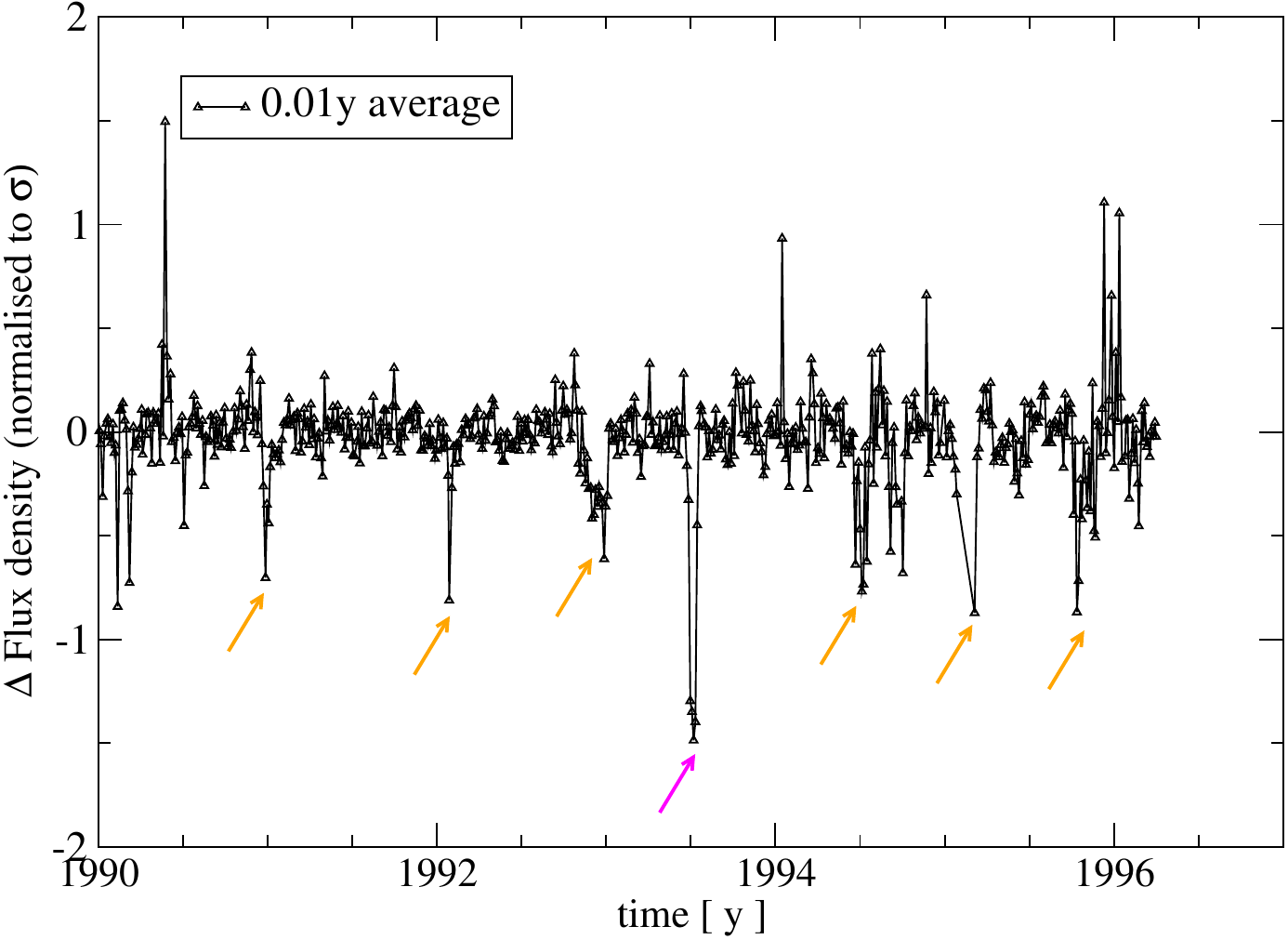}
      \caption{ The combined 2\,GHz variability curve of all the NESMP data: for each source, the flux densities have been de-trended, and normalised according their standard deviation $\sigma$; afterwards, the data of all the sources have been stacked into a single light curve, which has been averaged over time bins of 0.01\,y. A one-year periodic ripple, resulting from the combination of the annual oscillations in the light curves, has been removed from the stacked data. The regular sequence of dips, with average separation of 0.5\,y, is indicated with orange arrows, while a magenta arrow is reserved for the 1993.5 ESE.}
      \label{fig:allTime}
\end{figure}

We analysed the combined variability characteristics of all NESMP sources as a function of time to check whether more episodes of correlated variability, similar to the July 1993 event, can be found in the data. All the 2\,GHz light curves have been de-trended, and then normalised, imposing the same \emph{standard deviation} to all light curves, to ensure that the fast variability of all sources has a comparable weight in the calculation of the combined variability curve; finally, the data have been stacked. The resulting data-points have been averaged on intervals of 0.01\,y to generate an average combined variability curve. This turned out to be characterised by a smooth one-year periodic pattern, which results from the combination of the annual oscillations in the light curves, discussed in Paper II; after modelling and removing this systematic modulation, we obtained the combined variability curve in Fig. \ref{fig:allTime}. Episodes of time-correlated variability among many sources should result in clear variability features in the combined curve. The 1993.5 ESE is indeed recognisable as a short and sharp drop of the average flux (magenta arrow in the plot); six more events (orange arrows), all shaped approximately as an ESE, can be identified in the data, generally less intense but longer than the 1993.5 event. The temporal sequence of the events is the following: 1991.0, 1992.1, 1993.0, 1993.5, 1994.5, 1995.1, and 1995.9. The 1993.5 event appears to be an extreme manifestation of a long trail of similar phenomena with a cadence of $\sim6$ months, approximately falling at the end and in the middle of the year.

The semi-annual pattern is detectable, but without the same regularity, also in the 8 GHz data, which display ESE-shaped variability in 1990.5, 1993.0, 1994.5, and 1995.4. Before the year 1990 the periodic pattern, both at 2 and at 8 GHz, is less evident, as the variability in the combined variability curve is considerably stronger; most likely, this is due to the low number of sources monitored before the GBI upgrade (40 objects, less than a third of the later sample), which makes it harder to isolate correlated variability from the one that is intrinsic to the sources.

\section{SRV classification}
\label{sec:cls}

Through the analysis reported above we identified four distinct kinds of ESE-like variations in the data, which differ in duration, dependence (on solar elongation or time), and recurrence (which is the discriminating factor between type IIa and type IIb SRV). Their main properties are summarised in Table \ref{tab:types}; they all have in common the shape (resembling an ESE) and the achromaticity, although the importance of the effect at different frequencies changes not only according to the variability type, but, often, also from source to source.

All sources of the NESMP have been classified in Col. 9 and 10 of Table \ref{tab:allInfo} according to the kind of SRV found through a visual inspection of the 2 GHz and 8 GHz light curves, respectively. The reader should be warned that, while for some objects the classification is straightforward, for others it cannot be given without uncertainty. This is mainly because of the superposition of different effects, and, for a few sources, because of the limited amount of available data. 

Sources showing episodic variability (e.g. 0333+321) have been classified as affected by SRV. 
The presence of time-dependent variability, indicated in Col. 11, should be considered as purely indicative: TDV occurs in many sources in 1993.5; we classified a source as affected by TDV if at least one more event at 0.0 or 0.5\,y could be detected. There is an obvious ambiguity between TDV and type IIb SRV in sources whose minimum/maximum solar elongation occurs around 0.0 or 0.5\,y. The duration of the events can often remove the ambiguity (type IIb SRV is characterised by longer timescales); sources for which the two effects cannot be unambiguously identified are indicated as being affected by both.

\begin{table}
\caption{Main characteristics of the different types of systematic variability found in the data.\label{tab:types}. The acronym SE in the dependency (Dep.) column indicates solar elongation.}  
\centering
\begin{tabular}{l c c c c}
\hline
Type & Origin & Dep. & Duration & Periodicity \\
\hline
 I SRV & Instrumental & SE & weeks & annual \\
 IIa SRV & Unknown & SE & months & annual \\
 IIb SRV & Unknown & SE & months & semi-annual \\
 TDV & Unknown & Time & weeks & semi-annual \\
\hline
\hline
\end{tabular}
\end{table}

\subsection{Automated classification of sources}

Given some unavoidable degree of subjectivity in the classification of the sources according to a visual inspection, we developed a simple method for their automatic classification. The latter is not meant to replace the former, which is deeper and more accurate, but it provides an important tool to assess the reliability of its results.

It is based on the usage of NDCF, for the identification and localisation of dips in the light curves; the number of dips whose timing is consistent with SRV/TDV is compared with the null hypothesis that dips can randomly occur at any time of the year, independently of solar elongation.

\begin{figure}
   \centering
   \includegraphics[width=0.96\columnwidth]{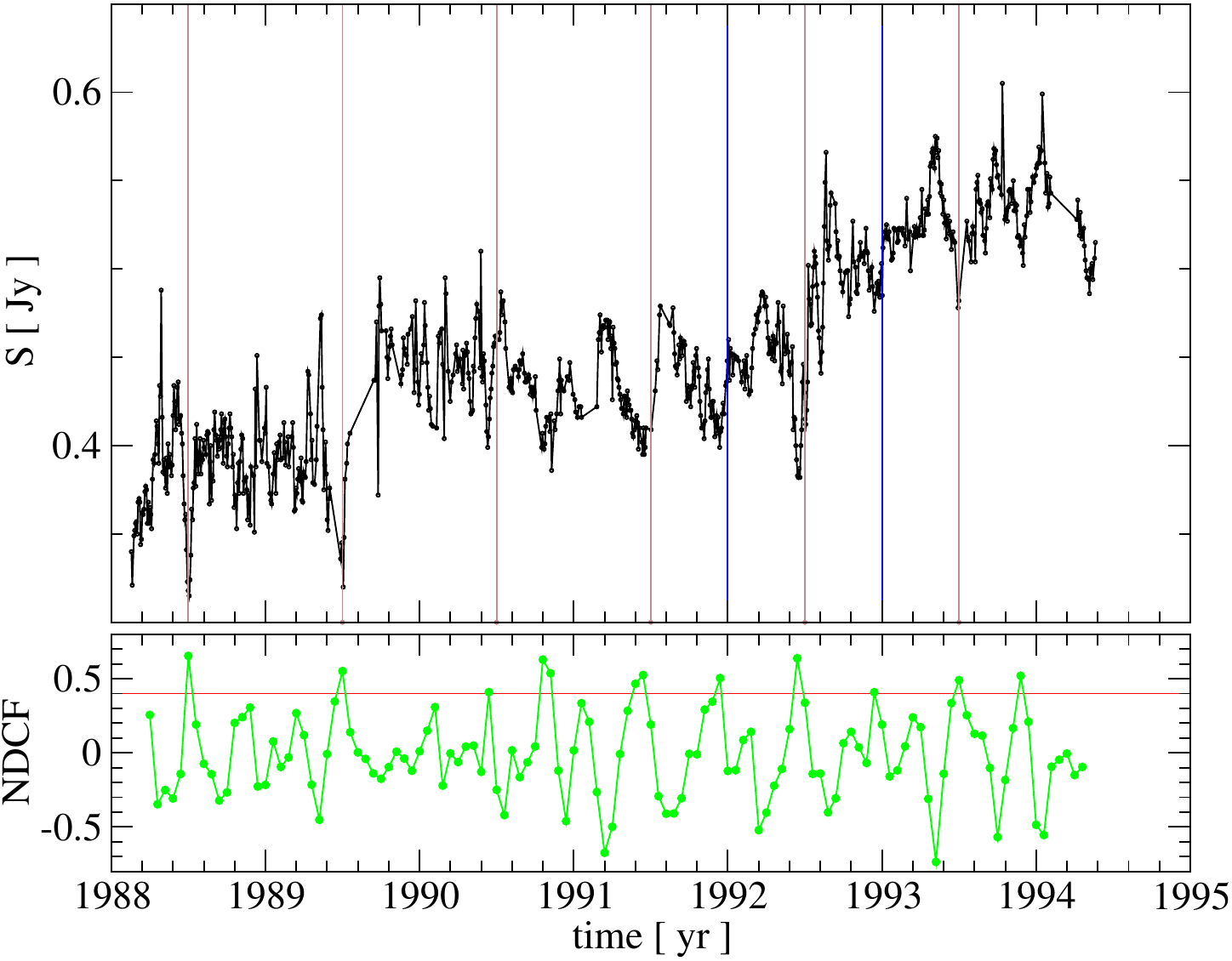}
      \caption{An example of the automatic detection of ESE-like dips in the light curves: the lower panel shows the NDCF output for the light curve of the type IIb-SRV source 1830+285 (plotted in the upper panel). Events above the threshold of 0.4 (red horizontal line) are regarded as detections. Blue and brown vertical lines indicate, respectively, the time of minimum and maximum solar elongation of the source.}
      \label{fig:ADex}
\end{figure}

More specifically, we created a generic dip model as a 0.2y V-shaped flux density drop, flanked by 0.1y-wide segments of constant flux. We cross-correlated the original light curve (i.e. without de-trending) of each source with the generic dip model. A high peak in the NDCF indicates the presence (and the location) of a feature in the light curve similar to the dip model. Peaks are considered significant if they are higher than a given threshold. For most sources, a degree of correlation of 0.4 produced a reasonable amount of events; when less than 2 or more 10 events were detected, the analysis has been repeated with a correlation threshold of 0.3 or 0.5, respectively. The choice of a variable threshold originates from the need to overcome two problems: a high level of noise in the data proportionally reduces the level of correlation between the light curve and the dip model, causing a decrease of detected events which can be compensated by a decrease of the threshold; a high level of source-intrinsic variability, on the other hand, causes spurious detections of dips, which requires a higher threshold to limit the number of false alarms. The detection of few or many events in a light curve with threshold 0.4 is a rather reliable indicator of the presence of either of the problems described above. By allowing the algorithm to overcome them through a variation of the threshold we can keep the procedure as automatic as possible. An example of how the algorithm works is provided in Fig. \ref{fig:ADex} for the type IIb SRV source 1830+285. In the lower panel, the peaks of the NDCF above the threshold correspond to detections of dips in the light curve (upper panel). It is worth to note the presence of negative dips in the NDCF, which correspond to rapid outbursts in the light curves. Since the algorithm we developed is very sensitive to sharp flux density variations, it can reveal, with opposite signs, both dips and flares.

The full list of identified events, for all sources, is reported in Appendix \ref{sec:AppDips}; those that are undoubtably caused by TDV are shown in boldface.

For type I SRV, events are expected to occur exactly at the time of minimum solar elongation. 
For type IIa and IIb, SRV dips are broader, and sometimes their time of occurrence can be determined less precisely; therefore, NDCF peaks occurring within 0.05y from the solar elongation minimum (or maximum) are still consistent with type IIa (or type IIb) SRV. To take into account these differences in the expected location of the dips, for each source we calculated four indicators: I. number of events occurring exactly at solar elongation minimum; II. number of events occurring within 0.05y from solar elongation minimum; III. number of events occurring exactly at solar elongation minimum or maximum; IV. number of events occurring within 0.05y from solar elongation minimum or maximum. Since the NDCF was calculated with a step of 0.05y, a chance occurrence of an event at solar elongation minimum for indicator I is 0.05; for indicator II it is 3x0.05=0.15; for indicator III 2x0.05=0.1; for indicator IV 6x0.05=0.3. 

Through a binomial probability distribution, we can calculate the probability that the number of successes (i.e. the events occurring at minimum/maximum solar elongation) is consistent with the null hypothesis that they happen at random time. If the probability is lower than 0.05 the source is classified as affected by SRV. If the lowest probability is returned for indicators I or II, the source is classified as type I or IIa SRV, otherwise as type IIb SRV. 

Subsequently, the same procedure has been applied three more times to the data: a first time after removing data-points obtained when the source's solar elongation is lower than 16\dgr; if the resulting light curve is still affected by SRV, it is classified as a type IIa, otherwise as a type I. A second time, after redefining successful detections as events occurring around day-of-the-year 0 (0.0y) or 182.5 (0.5y), in order to assess TDV.  Lastly, after redefining successful detections as events occurring 0.25y before minimum/maximum solar elongation: as there is no reason to expect dips to systematically appear at this time, this test provides an estimate of the false positives returned by the analysis.

\subsubsection{Results}

Despite its simplicity, the algorithm is effective. The original classification of sources is confirmed for 80\% of the sample (120 out of 148 sources): 83 out of 90 sources are confirmed as showing no SRV; 19 out of 26 sources are confirmed as type I SRV; 5 out of 7 as type IIa; 8 out of 12 as type IIb (see Col. 13 and 15 of Table \ref{tab:allInfo} for the automatic SRV and TDV classification; the probability of a random occurrence of the events is given in Col. 12 and 14). Among the uncertain cases, 4 out of 8 type IIa? and 1 out of 5 type IIb? are classified as type IIa and type IIb respectively. In total, with the automatic classification of sources we obtain 95 no-SRV sources, 19 type Ia SRV, 20 type IIa SRV, and 14 type IIb SRV (the classes of uncertain cases disappear). The group of 95 no-SRV sources includes also 11 objects for which too few data are available for a proper classification.

The search for sources highly affected by TDV returned 9 objects, 4 of which are also showing  type IIb SRV.
Quite remarkable is  the result provided by the algorithm when successes are defined as the events occurring 0.25y before minimum/maximum solar elongation: this way, only one object (i.e. 2234+282) out of 148 is identified as affected by  periodic dips, which shows that the number of false positive detections to be expected in the entire sample is very low.

The list of events found through automatic detection can also be used to identify the most important TDV events. We divided the timespan of GBI observations (1979-1996) in bins of 0.05y width; after flagging dips that are consistent with SRV (because of their time of occurrence and the classification of the source in which they were detected), we counted the number of dips in each bin. The bins with the highest number of counts are the more likely to be affect by TDV. By setting a threshold of 8 counts, we found the following sequence of TDV events: 1991.00, 1991.55, 1992.50, 1992.95-1993.00, 1993.50-.55. Note that before 1988 relatively few sources were monitored, so it is not possible to identify TDV events through this method. The sequence above does not coincide with the one obtained via the inspection of Fig. \ref{fig:allTime}; the semi-annual cadence of TDV, however, is confirmed.

There seems to be no obvious similarity between the sources showing simultaneous dips in the light curves. We identified 10 events around 1991.00y that, because of their shape and duration, appear to be certainly due to TDV; they affect the light curves of the following objects: 0316+413, 0337+319, 0400+258, 0723+679, 1123+264, 1150+812, 1538+149, 2007+776, 2032+107, 2105+420. The sources are widely spread both in right ascension and in declination, which seems to exclude both a common calibration scheme or a possible issue  related to the pointing towards a specific region of the sky.

\subsubsection{Caveat}

A complete automation of the procedure is hampered by the ambiguity between TDV and SRV for sources reaching minimum/maximum solar elongation around 0.0/0.5y. If the algorithm detects an event around these times, it is necessary to proceed with a visual inspection to properly classify it, although in some cases the ambiguity cannot be resolved. If the shape of the event is equal to the one seen in the combined light curve (see Fig. \ref{fig:allTime}) its classification as TDV is straightforward; if the shape/duration of the event is clearly different, the classification as SRV is obvious too. Sometimes, instead, the event's shape suggests a probable superposition of the two effects, and therefore a classification as TDV+SRV.

The most complicated case, in this respect, is the one of 0954+658. Out of the 7 events found by the automated detection algorithm, 4 occur either around 0.00 or 0.50y, while only one (the archetype of the ESEs) is at maximum solar elongation; this would suggest TDV as the origin of the variability, rather than SRV. The duration of the events, however, is not consistent with the features in the 2 GHz combined variability curve. A further complication comes from the very high degree of correlation between the fast variability of the source (obtained through a de-trending interpolation timescale of ten months) and the ones of the angularly-nearby sources 0633+734, 0723+679, and 0836+710, with time delays that are consistent with the different solar elongation patterns followed by these objects. The correlation between the variability of different sources would be consistent with TDV, but the time delay between these variations implies a solar-elongation-related effect. Finally, the archetype event of 1981.1 is not the only clear episode of sharp flux density drops at maximum solar elongation: even if not detected by the automatic algorithm, three more events can be seen (in 1989.10, 1991.10, and 1992.10), which explain the dips found in the annual pattern of the source (visible both at 2 and 8 GHz, see Fig. \ref{fig:0954}) and the visual classification of its variability as type IIb SRV.

Given the facts illustrated above, the most probable explanation for the fast variability of 0954+658 is that both TDV and SRV contribute to it. The similarities with the light curves of  0633+734, 0723+679, and 0836+710 suggest a common origin of the variability. Note that the complexity of such cases cannot be resolved through a fully automatic analysis of the data.

The algorithm has been applied also to the 8 GHz data. The results, however, in this case are much less reliable. The variety of shapes and widths of the dips caused by SRV/TDV makes it difficult to create a single archetype that could efficiently detect the majority of the events. Only 16 sources are automatically classified as affected by type II SRV, and they often do not coincide with the 23 objects identified through visual inspection.

\section{Discussion}
\label{sec:discussion}

Among SRV manifestations, only type I events can be satisfactorily explained so far (see Sect. \ref{sec:typeI}). In Sect. \ref{sec:OJ}, \ref{sec:0537}, and \ref{sec:time} we identified and described three more types of systematic variability in the data, namely type IIa SRV, IIb SRV, and TDV; there are ambiguities and exceptions in the general properties we outlined for each variability type, which do not facilitate the formulation of a robust hypothesis concerning the origin and the possible relationships among these effects. It seems reasonable to assume that they are not independent of each other, and that there is a common ground to which the events can be attributed. 

Some hints concerning the nature of the variability may come from the distribution in the sky of the affected sources; however, the indications resulting from the visual inspection and the automatic classification of the sources are not very consistent among each other. The distribution obtained with the first method (see Fig. \ref{fig:srcDistrV}) suggests that type IIa SRV mainly affects objects at low ecliptic latitude within the right ascension range from 04h to 09h (corresponding to solar elongation minima between 0.39 and 0.59\,y). This range largely overlaps with the one comprising most of type IIb sources (from 05h to 10h, corresponding to solar elongation minima between 0.45 and 0.63\,y); type IIa and type IIb could therefore be manifestations of the same phenomenon, essentially differentiated by the ecliptic latitude of the sources (lower for the former, higher for the latter). A second block of type IIb sources can be found in a narrow right ascension stripe about 12h apart from the first one, around right ascension 18h; it corresponds to solar elongation minima between 0.89 and 0.02\,y. 

The automatic detection algorithm returns a distribution of sources that is consistent with the one of the visual inspection for type IIb SRV. Type IIa SRV objects, instead, are more widely spread in the sky than what the visual inspection suggests. The two classification methods agree though on the facts that this kind of variability mostly concerns objects at low ecliptic latitude, and that there is a large cluster of them around R.A.: 8h, Dec: +20\dgr.

Some hypothesis as to the origin of the variability are shortly discussed below; particular attention is paid to assess a possible role of ISS.

\subsection{SRV as a sequence of ESEs}

\begin{figure}
   \centering
   \includegraphics[width=1.0\columnwidth]{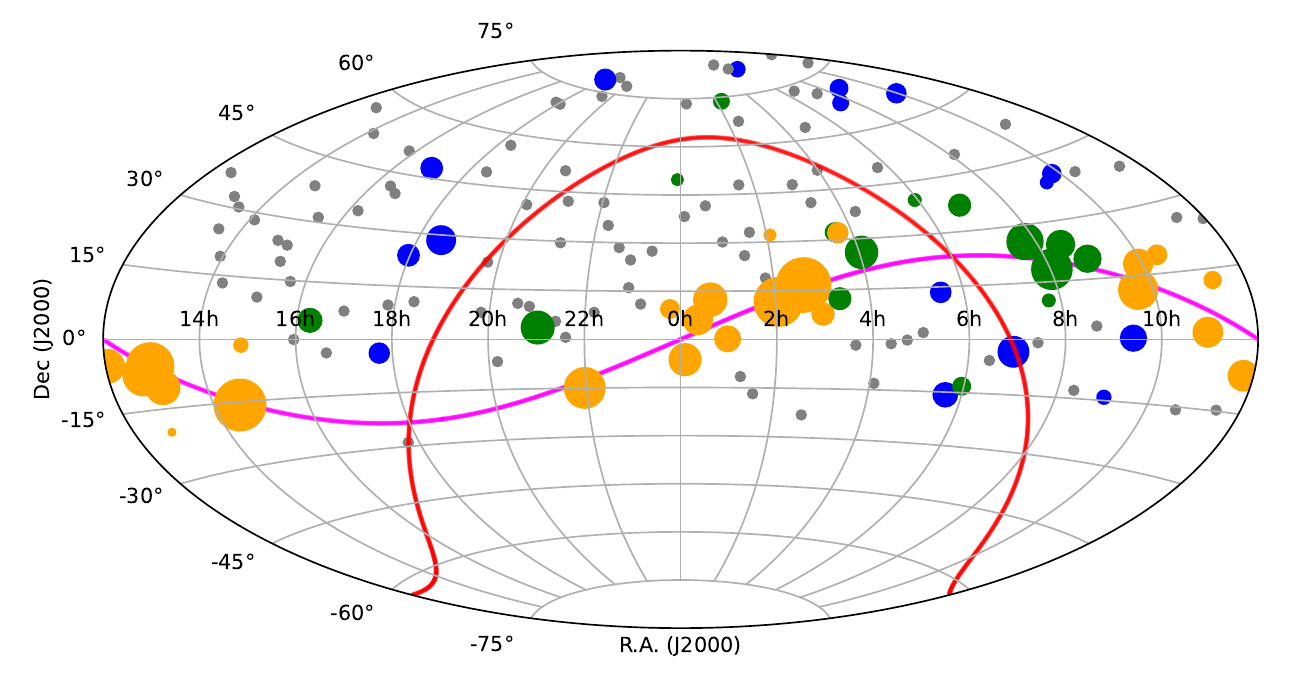}
      \caption{Visual inspection of sources: the distribution in the sky of type I (orange dots), type IIa (green dots), type IIb (blue dots), and unaffected sources (grey dots). The size of the dots is proportional to the ratio between the 2\,GHz dropout at minimum solar elongation and the average 2\,GHz flux density (Col. 8 and 4, respectively, in Table \ref{tab:allInfo}). The magenta line shows the ecliptic, while the red one shows the galactic plane.}
      \label{fig:srcDistrV}

   \centering
   \vspace{0.3cm}\includegraphics[width=1.0\columnwidth]{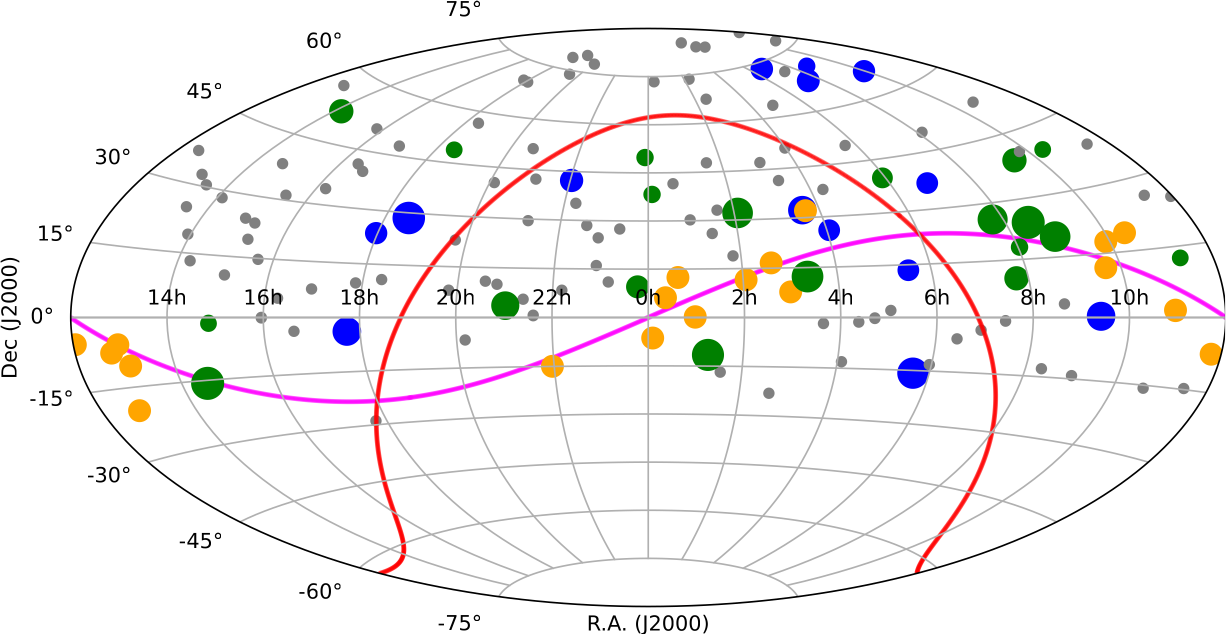}
      \caption{Same as Fig. \ref{fig:srcDistrV}, but for the automatic classification of sources. The size of the dots is proportional to the exponential of -(10p), where p is the false alarm probability returned by the algorithm (Col. 12 in Table \ref{tab:allInfo}).}
      \label{fig:srcDistrA}
\end{figure}

Since ESEs are generally attributed to ISS, it is important to assess whether a series of events that look like ESEs can be reasonably explained in terms of ISS too. Concerning TDV the answer is certainly no, as the simultaneity of the events in different sources excludes the intervention of a localised screen as required in ESEs. Concerning type II SRV, however, the reply is less obvious. In principle, a localised screen moving very slowly could cause repeated events occurring approximately at the same time of the year, every time that the line of sight (LOS) to the source crosses the screen again. The fact that, as shown in Fig. \ref{fig:srcDistrV} for the results of the visual inspection, sources affected by SRV are mainly concentrated near R.A. 6 and 18h would agree with this picture, as these are the conditions for which the ISM velocity transverse to the LOS is minimum, so the Earth’s orbital velocity is dominant. The search for events repeating regularly with semi-annual cadence could facilitate the detection of ESEs that occur around the time of minimum/maximum solar elongation of a source, because at larger angular distances the cadence would generally be more asymmetric (e.g. there could be 4 months and then 8 months intervals between consecutive events).

\begin{figure}
   \centering
   \includegraphics[width=0.85\columnwidth]{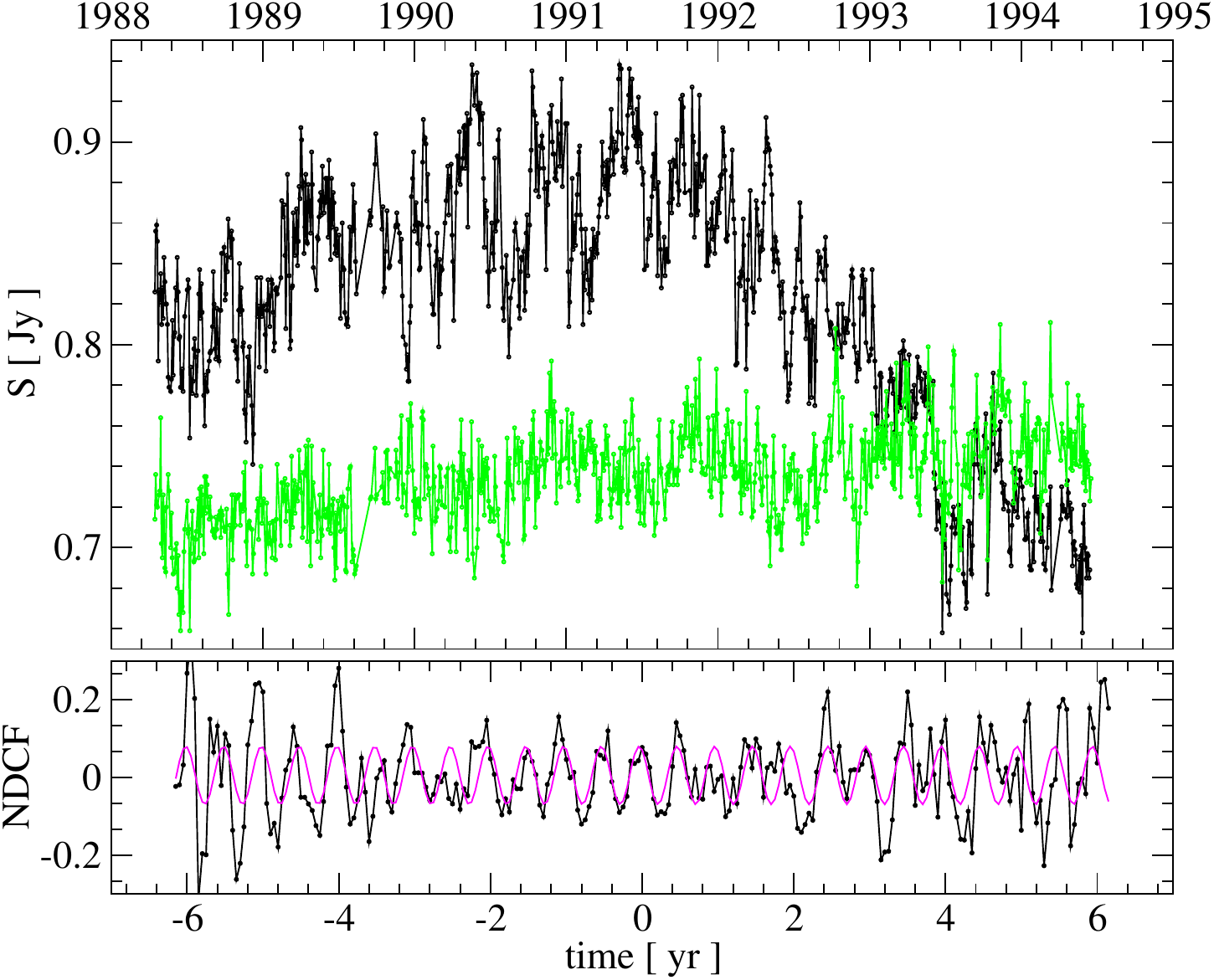}
      \caption{Upper panel: the light curves of the strong type IIb source 0922+005 (black dots) and of 0837+035 (green dots). The NDCF between the de-trended light curves is shown in the lower panel; it reveals a 0.5y periodic modulation (magenta line) with a time delay that is consistent with the difference in the solar elongation minima of the objects.}
      \label{fig:0922}
\end{figure}

There are, however, a number of critical arguments against this hypothesis:
\begin{itemize}
\item ISS fails to explain both TDV and type IIa SRV. If the 
dips at minimum solar elongation would be caused by yearly crossings of the same screen, dips would be found at maximum solar elongation too, as the Earth would align along the same LOS, therefore type IIa SRV would not exist.
\item The procedure for the automatic classification of type IIb SRV sources is the only one to be biased because of the requirement of a semi-annual cadence of the events. The automatic {\it detection} of events, the visual and automatic classification of type IIa SRV sources, as well as the visual classification of type IIb SRV objects are unbiased. They all agree upon the fact that dips are not randomly occurring across the year, but they are much more common at the time of minimum and maximum solar elongation, which cannot be explained without a direct correlation between SRV and the position of the Sun.
\item Events in the light curves of SRV sources often do not occur regularly every 6 or 12 months, but there can be large gaps between them. For example, in 0528+134, SRV events are found in 1990.50, 1991.00, 1992.40, and 1993.95. This is unexplainable, under the assumption that an almost motionless screen is causing the variability.
\item The case of 0954+658 and the angularly-nearby sources 0633+734, 0723+679, and 0836+710, which show correlated variability with time delays consistent with the time difference among the solar elongation minima, shows that SRV events can affect multiple sources in the same region of the sky. Another interesting example is provided by 0922+005, and the angularly-nearby source 0837+035 (see Fig. \ref{fig:0922}). The former is a strong type II SRV source; the variability of the latter is dominated by random noise, and, as a consequence, is not identified as affected by SRV. Nevertheless, the fast variability of the two objects is correlated: an NDCF analysis of their de-trended light curves shows regular peaks of correlation with 0.5y cadence, and a time delay of 0.04y, in excellent agreement with the difference in their solar elongation minima. A further example is provided by the quasi simultaneous events in OJ287 and 0742+103 (see Fig. \ref{fig:0742-OJ}). They all suggest that SRV is not caused by a very localised screen at large distance from the observer, but by a nearby, angularly-wide screen related to the position of the Sun.
\end{itemize}

A possible source of bias in favour of events occurring more often around the times of minimum/maximum solar elongation could be the higher transversal speed of the Earth with respect to the LOS, which implies that a larger region in the sky is swept around these times. This argument, again, cannot explain TDV, neither type IIa SRV, because, if the transversal speed would be the cause of SRV, the events would have the same chance to occur at minimum and at maximum solar elongation. Type IIb SRV, instead, would be compatible with an effect of the transversal speed; but this would become more important as the ecliptic latitude of the source tends to zero. Instead type IIb SRV concerns mostly sources at high ecliptic latitude where the effect would be marginal, while type IIa SRV seems stronger for sources at low ecliptic latitude. Finally, an important contribution of the transversal speed to SRV would necessarily lead to an equally important decrease of the duration of the events at the times of minimum/maximum solar elongation; the data, however, show absolutely no sign of it.

\subsection{Further hypotheses}

An instrumental origin of the variability can be ruled out for different reasons: all the main examples of the different kinds of variability (OJ287 for type IIa, 0528+134 for type IIb and TDV) are supported by observations from different facilities, whose flux density measurements are in excellent agreement with each other; the variability also affects night time observations (for type IIb SRV in particular), which excludes sunlight reflection as a possible cause; it is also worth mentioning that, differently from the instrumental type I SRV, the centre and the duration of type II and TDV events in the light curves changes from year to year, which means that the variability is not as systematic as one would expect from an instrumental effect.

Assuming that the source distribution returned by the visual inspection of light curves is correct, type II SRV mostly affects sources whose solar elongation minima and maxima occur around 0.0 and 0.5\,y; these are also the critical times for the appearance of TDV: time could therefore be the link between the different kinds of ESE-shaped events we detected in the data. Several astronomically-relevant effects or events occur around 0.0 and 0.5\,y, namely meteorological effects, perihelion$/$aphelion, the Earth's intersection of the Sun's equatorial plane, and solstices. We separately discuss them, briefly, in the context of systematic flux density variability, here below:

\begin{itemize}
\item {\bf meteorological effects}, such as storms, or cold/heat waves, are local (so they are not expected to affect in the same way different instruments), non-periodic, and cannot account for the solar-elongation dependence of SRV. Their involvement in SRV can safely be excluded.
\item The occurrence of {\bf perihelion and aphelion} seems unlikely to explain ESE-shaped variability, essentially because of the smoothness and the extent of the Sun's distance variation; also, it would be hard to understand how the Earth reaching the minimum and the maximum distance from the Sun could lead to the same consequences, as required by type IIb SRV.
\item The {\bf Earth's intersection of the Sun's equatorial plane} raises interest because of its 6-month periodic cadence; however, the angle between the ecliptic and the Sun's equatorial plane is small ($\sim7$\dgr) and the variation smooth. In order to produce the observed ESE-like features, a sharp change of the environmental characteristics at the crossing of the equatorial plane would be needed, which though does not seem to be supported by external evidence.
\item An interpretation of the variability as a consequence of the {\bf solstices} seems the most promising, although it does not come without problems. On the one hand, it is reasonable to consider that the variation of the Sun's declination during the year could have an impact on flux density measurements, for instance by affecting the propagation of radio waves through the atmosphere. On the other hand, there is no obvious reason why a source should be similarly affected when the Sun is at maximum and at minimum declination, as for type IIb SRV. 
\end{itemize}

The mechanism that could be responsible for the variability is equally hard to identify. It is important to preliminarily underline that the timescales and the amplitudes of the variations detected in the light curves are not consistent with the theoretical values from scattering in a nearby medium (see \citealt{1992RSPTA.341..151N} for an introduction to the physics of scintillation). 

The plausible existence of a ring that crosses the whole sky along the right ascension lines around 06h and 18h, comprising most of the type II SRV sources, if confirmed, would not be consistent with IPS because of its wide extension.
Since type II SRV affects the sources months before/after the time of minimum elongation, when the Sun's declination is far from its highest northern/southern extension, atmospheric scattering seems unlikely too.

The limited range of declination of type II sources could suggest a relationship with the Earth's North-South direction, implying a link to the Earth's geometry and properties. 
The fact that, within the 06h-18h right ascension ring, some sources are affected for a large part of the year, would indicate a geometrical effect as the most plausible. The plane on which the ring lies corresponds to the one defined by the equatorial and the ecliptic latitudes axes.
At a level of conjecture, one could wonder whether the complex shape of the magnetosphere could play a role in the observed variability, acting somehow as a lens. The magnetosphere stretches along the direction of the flow of solar wind, in the ecliptic plane; it is compressed towards the Sun, and very elongated in the opposite direction (magnetotail); its shape varies during the year, and its structure along the line of sight to a source depends on the source's position with respect to the Sun. The inner structure of the magnetosphere is also heavily influenced, because of the magnetic field, by the celestial poles. 
This implies the existence, in the morphology of the magnetosphere, of a privileged direction (the 6h-to-18h right ascension direction), towards the tilt of the Earth axis, which at 0.0 and 0.5\,y, approximately, aligns with the flow of the solar wind; such alignment implies a symmetric configuration of the magnetosphere with respect to Earth's axis that could possibly trigger both SRV and TDV. When the effect of the Earth axis modifies more strongly the inner structure of the magnetosphere, TDV would become dominant; when the effect is weaker, only sources with solar elongation minima around  0.0 and 0.5\,y would be affected, and we would see type IIb SRV. Although this scenario may be geometrically plausible, it has to face the fact that, to cause significant variability in the data, a lensing effect would require a change in refraction index that is orders of magnitude larger than what is expected at the level of all the atmospheric layers.

\section{Revision of identified ESEs}
\label{sec:eses}

The analysis of the NESMP data by \cite{2001ApJS..136..265L} led to the detection of 24 ESEs (see their Tables 5 and 6). We have already shown that some of them are caused by SRV. In Table \ref{tab:eses} we summarise our new classification of these events in the light of the SRV and TDV characteristic previously illustrated. A short explanation of the reasons behind this classification is reported in Appendix \ref{sec:App2}.

\begin{table}
\caption{Summary of the analysis of ES-like events reported by \citealt{2001ApJS..136..265L}. Col. 1 and 2 report the name of the source and the time of occurrence of the event. The plausible origin of the variability is reported in Col. 3.\label{tab:eses}}  
\centering
\begin{tabular}{l c c}
\hline
Source & Time & Origin \\
            & (year) & \\
\hline
0133+476 & 1988.2 & GBI transition \\
0133+476 & 1990.2 & ESE?   \\
0201+113 & 1991.3 & Type I SRV  \\
0202+319 & 1989.5 & TDV \\
0300+470 & 1988.2 & GBI transition \\
0300+470 & 1988.3 & ESE? \\
0333+321 & 1986.3 & Type II SRV \\
0333+321 & 1987.9 & ESE \\
0528+134 & 1991.0 & Type IIb SRV \\
0528+134 & 1993.5 & Type IIb SRV or TDV\\
0528+134 & 1994.0 & Type IIb SRV \\
0952+179 & 1992.6 & Type I SRV \\
0952+179 & 1993.6 & Type I SRV \\
0954+658 & 1981.1 & Type IIb SRV \\
1438+385 & 1993.5 & TDV \\
1502+106 & 1979.5 & TDV? \\
1502+106 & 1986.0 & ? \\
1611+343 & 1985.4 & ESE \\
1741-038 & 1992.5 & Type IIb SRV \\
1749+096 & 1988.1 & GBI transition \\
1756+237 & 1993.7 & ESE? \\
1821+107 & 1984.2 & ESE \\
2251+244 & 1988.9 & TDV \\
2352+495 & $\sim$1985.0 & ? \\
\hline
\end{tabular}
\end{table}

The number of ESEs that cannot be explained through SRV is very low. Even the archetype of ESEs, the 1981.1 event in 0954+658, seems more likely to be caused by the Sun than by an interstellar screen. This means that the actual frequency of ESEs in the GBI data is significantly lower than previously claimed. Furthermore, this result demonstrates the difficulty to identify ESEs uniquely from the examination of light curves; for a robust detection, it would be necessary to use independent diagnostic tools confirming the interstellar medium as the origin of the variability.

\section{Conclusions}
\label{sec:concl}

We reported about a variety of events, shaped as ESEs, detected in the radio light curves of compact extragalactic objects, from GBI, UMRAO, Pico Veleta, and Effelsberg observations. We distinguished between four different kinds of variations, three of which are centred at the time of minimum solar elongation, while the fourth depends on time. Type I SRV has an instrumental origin, with a possible further contribution from IPS; 
it causes dips in the light curves when sources are at solar elongation below 15-20\dgr at 2 GHz (<$\sim$10\dgr at 8 GHz), with average flux density decreases of about 30 percent. Differently from type I, type II SRV can have a strong impact on the light curves even at high solar elongation. We found hints, which though need confirmation, that it mainly affects sources in right ascension ranges around 06h and 18h. The origin of the variability is unknown, but the detection of the effects in data from different facilities rules out an instrumental problem. TDV is a semi-annual effect affecting the light curves of a large number of sources around 0.0 and 0.5\,y; its nature, as for type II SRV, is yet unknown. The evidence we collected rules out both ISS and instrumental/calibration problems from the possible sources of the variability. By exclusion, the most viable option seems to be a propagation effect in a local screen (the IPM or the atmosphere).

After analysing the general properties of the four kinds of variability, we assessed different hypotheses about their origin; 
all hypotheses are weakened by critical arguments that cannot be overcome without assuming that the actual properties of local screens substantially differ from the ones proposed by current models. It cannot be ruled out that the variability is caused by a superposition of different effects, which could explain the difficulty to describe it in terms of a single self-consistent model. However, the convergence of many SRV and TDV manifestations around the time of the solstices, if confirmed, would strongly support the idea of a common origin.

From the picture sketched above, it is evident that  the nature of the variability is far from being identified. Much clearer is the situation concerning the consequences of SRV/TDV, especially in relation to ESEs studies.
Many of the ESEs identified by \cite{2001ApJS..136..265L} during the NESMP (which is the most important program dedicated to the detection of these events) are certainly due to SRV; only a few are not compatible with the characteristics of a solar influence. Even the impressive events in the light curves of 1741-038 (see \citealt{2001ApJ...546..267L}), 0954+658 (\citealt{1987Natur.326..675F}; \citealt{1987ApJS...65..319F}; \citealt{1998ApJ...498L.125W}), and 0528+134 (\citealt{1996A&AS..120C.529P}) appear to be extreme cases of semi-annual events caused by the Sun (type IIb SRV). In the light of the discoveries here reported, also the ESE detected in the data of PKS 1939-315 on June 2014 (\citealt{2016Sci...351..354B}) would be naturally interpreted as due to SRV or TDV, considering that, because of its coordinates, the source would be expected to show type IIb SRV, and that the minimum of the symmetric event falls at the beginning of July 2014 (i.e. the critical time 0.5\,y), when the source reaches its maximum solar elongation. Our study does not rule out the existence of ESEs, but it shows that this kind of event, which already \cite{1994ApJ...430..581F} considered rare, is even rarer than previously thought.

Concerning the study of the radio properties of compact sources, the existence of SRV implies that the variability characteristics of some objects may be heavily affected by the influence of the Sun. For sources such as OJ\,287, it may be important to set a much larger Sun constraint than the mechanical one allowed by the telescope (which for Effelsberg, e.g., is 2\dgr, see \citealt{2023ApJ...944..177K}). Particularly worrisome is the recognition of the high degree of correlation between the fast variability of 0954+658 and of some angularly-nearby sources, whose solar elongation is always high.  Periodicity in light curves with periods close to one year or 6 months should be regarded as highly suspicious, especially if they appear in sources whose solar elongation peaks around the critical times 0.0 and 0.5\,y; in this respect, it would be interesting to check whether the $\sim$180-day periodicity found in OVRO data for the NLSy1 J0849+5108 (\ecl: 32\dgr; see \citealt{2021ApJ...914....1Z}), whose coordinates are compatible with type IIb SRV, shows dips around 0.1 and 0.6\,y, consistently with its time of maximum/minimum solar elongation. Additional studies will be needed to assess whether SRV may also play a role in the recently discovered symmetric achromatic variability events (SAV; see \citealt{2017ApJ...845...89V}), found in the blazar 1413+135; however, the position of the source (far from the location of most type IIa and IIb SRV sources), the duration of the events (of the order of one year or more), and the occurrence of the detected SAV events (see \citealt{2022ApJ...927...24P}) all seem to indicate that the variability is not caused by the Sun.

A thorough investigation of the causes and the manifestations of SRV is made even more urgent by the consideration that several of the most variable compact radio sources of extragalactic origin (e.g. OJ\,287 and 0235+164) fall at low ecliptic latitudes, and that Sgr A*, among the most interesting and studied objects in astronomy, because of its coordinates and an ecliptic latitude of only $-5.6$\dgr, should be particularly sensitive to type II SRV.

\clearpage
\onecolumn
\begin{landscape}
\begin{longtable}{lrcccccc|ccc|rlrl}
\caption{\label{tab:allInfo} The main variability characteristics of NESMP sources are reported as follows: name of the source (Col. 1), ecliptic latitude (Col. 2), time of the year in which solar elongation reaches the minimum (Col. 3), average S$_{2\mathrm{GHz}}$ (Col. 4), S$_{2\mathrm{GHz}}$ standard deviation (Col. 5), average S$_{8\mathrm{GHz}}$ (Col. 6), S$_{8\mathrm{GHz}}$ standard deviation (Col. 7), 2\,GHz dropout at minimum solar elongation (Col. 8), SRV classification at 2 GHz (Col. 9), SRV classification at 8 GHz (Col. 10), evidence of TDV (Col. 11). The results of the automatic classification of sources according to their 2 GHz light curves are summarised in Col. 12-15; they report the probability that the number of SIV or TDV events in the light curves is consistent with an homogeneous distribution of dips across the year (Col 12 and 14, respectively), and the consequent classification of the sources (Col. 13 and 15). The "NA" indication is used when no meaningful estimate could be achieved because of lack of data.}\\
\hline\hline
 Source  &  $\beta$ & SE & $\overline{S}_{2\mathrm{GHz}}$ & $\sigma_{2\mathrm{GHz}}$ & $\overline{S}_{8\mathrm{GHz}}$ & $\sigma_{8\mathrm{GHz}}$ & 2GHz  & \multicolumn{3}{c}{Visual Classification}  & \multicolumn{4}{c}{Automatic Classification (2 GHz)} \\
         &          &    minimum    &                     &   &                     &                        &  dip &  \multicolumn{2}{c}{SRV}  & TDV & Prob & SRV & Prob & TDV\\
         &  (\dgr)  & (yr)    &  (Jy)               &   (Jy)                  &      (Jy)           &    (Jy)                & (Jy) & 2 GHz & 8 GHz & & & & & \\
\hline
\endfirsthead
\caption{continued.}\\
\hline\hline
 Source  &  $\beta$ & SE & $\overline{S}_{2\mathrm{GHz}}$ & $\sigma_{2\mathrm{GHz}}$ & $\overline{S}_{8\mathrm{GHz}}$ & $\sigma_{8\mathrm{GHz}}$ & 2GHz  & \multicolumn{3}{c}{Visual Classification}  & \multicolumn{4}{c}{Automatic Classification (2 GHz)} \\
         &          &    minimum    &                     &   &                     &                        &  dip &  \multicolumn{2}{c}{SRV} & TDV & Prob & SRV & Prob & TDV\\
         &  (\dgr)  & (yr)    &  (Jy)               &   (Jy)                  &      (Jy)           &    (Jy)                & (Jy) & 2 GHz  & 8 GHz & & & & & \\
\hline
\endhead
\hline
\endfoot
0003+380  &  34.1  &   0.27   &   0.59   &   0.04  &   0.90  &   0.19  &  0.000  &   &  & & 0.047 & IIa & & \\
0003-066  &  -6.5  &   0.22   &   1.98   &   0.28  &   2.22  &   0.59  &  0.220$\pm0.03$  & I &  & & 0.0012 & I & & \\
0016+731  &  60.4  &   0.37   &   1.39   &   0.16  &   1.64  &   0.20  &  0.000  &   &  & & & & & \\
0019+058  &   3.4  &   0.24   &   0.43   &   0.08  &   0.45  &   0.21  &  0.043$\pm0.012$  & I & I  & & 0.0034 & I & & \\
0035+121  &   7.7  &   0.26   &   0.75   &   0.02  &   0.38  &   0.03  &  0.093$\pm0.019$  & I &  & & 0.00000 & I & & \\
0035+413  &  34.0  &   0.29   &   0.97   &   0.07  &   1.01  &   0.12  &  0.000  &   &  & & & & & \\
0055+300  &  22.2  &   0.29   &   0.42   &   0.01  &   0.68  &   0.07  &  0.000  &   &  & & & & & \\
0056-001  &  -5.7  &   0.25   &   1.95   &   0.05  &   0.87  &   0.08  &  0.140$\pm0.014$  & I &  & & 0.00000 & I & & \\
0113-118  & -18.2  &   0.25   &   1.62   &   0.17  &   1.49  &   0.24  &  0.000  &   &  & & 0.0025 & IIa & & \\
0123+257  &  15.7  &   0.30   &   1.14   &   0.03  &   0.96  &   0.09  &  NA  &   &  & & & & & \\
0130-171  & -24.6  &   0.26   &   0.77   &   0.02  &   0.60  &   0.07  &  NA  &   &  & & & & & \\
0133+476  &  34.8  &   0.33   &   1.51   &   0.17  &   1.99  &   0.45  &  0.000  &   &  & & & & & \\
0134+329  &  21.4  &   0.31   &  10.92   &   0.24  &   3.34  &   0.21  &  0.000  &   &  & & & & & \\
0147+187  &   7.2  &   0.31   &   0.41   &   0.04  &   0.45  &   0.08  &  0.000  &   &  & & & & & \\
0201+113  &  -0.9  &   0.31   &   0.91   &   0.04  &   0.73  &   0.10  &  0.24$\pm0.02$  & I & I & Y & 0.00000 & I & & \\
0202+319  &  18.3  &   0.33   &   0.89   &   0.05  &   0.87  &   0.17  &  0.013$\pm0.006$  & I &  & Y & 0.0059 & IIa & & \\
0212+735  &  54.9  &   0.40   &   2.38   &   0.09  &   2.92  &   0.58  &  0.06$\pm0.02$  & IIa? & & Y & & & &  \\
0224+671  &  48.9  &   0.39   &   1.41   &   0.24  &   1.68  &   0.48  &  0.000  &   &  & & & & & \\
0235+164  &   1.1  &   0.34   &   1.81   &   0.68  &   1.99  &   0.87  &  0.6$\pm0.2$  & I & I & Y & 0.0038 & I & & \\
0237-233  & -36.6  &   0.30   &   4.82   &   0.09  &   2.00  &   0.32  &  0.000  &   &  & & & & & \\
0256+075  &  -8.8  &   0.34   &   0.76   &   0.09  &   0.61  &   0.12  &  0.041$\pm0.010$  & I & I & Y & 0.023 & I & & \\
0300+470  &  28.7  &   0.38   &   1.58   &   0.29  &   1.76  &   0.57  &  0.000  &   &  & Y & & & & \\
0316+413  &  22.3  &   0.38   &  34.22   &   1.79  &  30.41  &   4.16  &  0.000  &   &  & & & & & \\
0319+121  &  -5.9  &   0.36   &   1.56   &   0.08  &   1.05  &   0.12  &  0.08$\pm0.02$  & IIa? & IIa? & & 0.003 & IIa & & \\
0333+321  &  12.6  &   0.39   &   2.66   &   0.11  &   1.57  &   0.20  &  0.10$\pm0.04$  & IIa & IIa & & 0.011 & IIb & & \\
0336-019  & -20.7  &   0.36   &   2.37   &   0.20  &   2.34  &   0.36  &  0.000  &   &   & & & & & \\
0337+319  &  12.2  &   0.39   &   0.23   &   0.01  &   0.03  &   0.01  &  0.009$\pm0.003$  & I &   & Y & 0.023 & I & & \\
0355+508  &  29.7  &   0.41   &   6.10   &   0.21  &   6.07  &   2.95  &  0.000  &   &   & & & & & \\
0400+258  &   5.2  &   0.40   &   0.97   &   0.05  &   0.77  &   0.05  &  0.11$\pm0.03$  & IIa & IIa & Y & 0.031 & IIb & 0.031 & Y \\
0403-132  & -33.3  &   0.37   &   3.24   &   0.10  &   1.65  &   0.16  &  0.000  &   &   & & & & & \\
0415+379  &  16.4  &   0.41   &   0.63   &   0.04  &   2.22  &   0.24  &  NA  &   &   & & & & & \\
0420-014  & -22.6  &   0.40   &   3.60   &   0.34  &   4.20  &   1.04  &  0.000  &   &   & & & & & \\
0440-003  & -22.3  &   0.41   &   1.54   &   0.12  &   0.94  &   0.20  &  0.000  &   &   & Y & & & & \\
0444+634  &  40.7  &   0.44   &   0.43   &   0.07  &   0.59  &   0.15  &  0.000  &   &   & & & & & \\
0454+844  &  61.2  &   0.47   &   0.42   &   0.04  &   0.43  &   0.06  &  0.000  &   &   & & & & & \\
0500+019  & -20.6  &   0.43   &   2.42   &   0.07  &   1.46  &   0.11  &  0.000  &   &   & Y & & & & \\
0528+134  &  -9.7  &   0.45   &   2.27   &   0.32  &   3.50  &   2.01  &  0.10$\pm0.03$  & IIb & IIb & Y & 0.031 & IIb & 0.0038 & Y \\
0532+826  &  59.2  &   0.47   &   0.26   &   0.01  &   0.22  &   0.04  &  0.000  &   &   & & & & & \\
0537-158  & -39.2  &   0.45   &   0.47   &   0.02  &   0.30  &   0.04  &  0.030$\pm0.008$  & IIb & IIa &  & 0.004 & IIb & & \\
0538+498  &  26.5  &   0.46   &  14.99   &   0.29  &   4.90  &   0.34  &  0.000  &   &   & & & & & \\
0552+398  &  16.4  &   0.47   &   3.67   &   0.10  &   5.79  &   1.17  &  0.06$\pm0.02$  & IIa? & IIb? & & 0.034 & IIa & & \\
0555-132  & -36.7  &   0.47   &   0.76   &   0.03  &   0.64  &   0.12  &  0.025$\pm0.010$  & IIa? & IIa? & Y & & & & \\
0615+820  &  58.6  &   0.48   &   0.83   &   0.03  &   0.67  &   0.06  &  0.020$\pm0.008$  & IIb? & IIa? & Y & & & & \\
0624-058  & -29.1  &   0.50   &   9.40   &   0.17  &   2.67  &   0.20  &  0.000  &   &  & & & & & \\
0633+734  &  50.1  &   0.48   &   0.92   &   0.03  &   0.83  &   0.08  &  0.000  &   &  & & 0.028 & IIb & & \\
0650+371  &  14.2  &   0.50   &   1.03   &   0.10  &   0.83  &   0.10  &  0.053$\pm0.013$  & IIa? &  & Y & 0.031 & IIb & & \\
0653-033  & -26.1  &   0.52   &   0.49   &   0.10  &   0.52  &   0.12  &  0.050$\pm0.014$ & IIb? & IIb? & Y & & & & \\
0716+714  &  48.6  &   0.50   &   0.45   &   0.05  &   0.62  &   0.18  &  0.000  &   &  & & & & & \\
0723+679  &  45.3  &   0.50   &   0.63   &   0.03  &   0.36  &   0.03  &  0.016$\pm0.004$  & IIb? &  & Y & 0.027 & IIb & 0.0081 & Y \\
0723-008  & -22.6  &   0.54   &   1.54   &   0.08  &   1.20  &   0.16  &  0.000  &   &  & & & & & \\
0742+103  & -10.9  &   0.55   &   4.29   &   0.09  &   2.84  &   0.30  &  0.070$\pm0.018$  & IIa & IIa & & 0.023 & IIa & & \\
0743+259  &   4.6  &   0.54   &   0.49   &   0.04  &   0.39  &   0.13  &  0.070$\pm0.010$  & IIa & IIa & & 0.0073 & IIa & & \\
0759+183  &  -2.2  &   0.56   &   0.50   &   0.04  &   0.56  &   0.07  &  0.09$\pm0.02$  & IIa? & & & 0.047 & IIa & & \\
0804+499  &  29.0  &   0.54   &   1.10   &   0.16  &   1.26  &   0.37  &  0.000  &   &  & & & & & \\
0818-128  & -31.5  &   0.59   &   0.85   &   0.08  &   0.60  &   0.09  &  0.000  &   &  & & & & & \\
0827+243  &   5.1  &   0.57   &   0.65   &   0.06  &   0.79  &   0.30  &  0.056$\pm0.015$  & IIa & IIb & & 0.0005 & IIa & & \\
0836+710  &  50.2  &   0.53   &   3.09   &   0.23  &   1.83  &   0.26  &  0.10$\pm0.03$  & IIb & IIb & Y & 0.047 & IIb & &  \\
0837+035  & -14.5  &   0.59   &   0.73   &   0.02  &   0.56  &   0.05  &  0.000  &   &  & & & & & \\
0851+202  &   2.6  &   0.59   &   2.52   &   0.30  &   3.82  &   1.05  &  0.20 $\pm0.07$ & IIa & IIa & & 0.0059 & IIa & & \\
0859-140  & -29.7  &   0.62   &   2.49   &   0.05  &   1.51  &   0.11  &  0.050$\pm0.010$  & IIb & & Y & & & & \\
0922+005  & -14.1  &   0.63   &   0.82   &   0.07  &   0.53  &   0.08  &  0.060$\pm0.007$  & IIb & & Y & 0.011 & IIb & & \\
0923+392  &  22.7  &   0.59   &   4.92   &   0.14  &   7.76  &   2.97  &  0.08$\pm0.04$  & IIb & IIb & Y & 0.023 & IIa & & \\
0938+119  &  -2.0  &   0.63   &   0.18   &   0.01  &   0.08  &   0.02  &  0.030$\pm0.005$  & I & I & & 0.00000 & I & & \\
0945+408  &  25.8  &   0.60   &   1.42   &   0.13  &   1.61  &   0.39  &  0.049$\pm0.010$  & IIb & IIb & Y & & & 0.028 & Y \\
0952+179  &   4.7  &   0.63   &   0.99   &   0.03  &   0.54  &   0.06  &  0.093$\pm0.011$  & I & I & Y & 0.00000 & I & & \\
0954+658  &  48.8  &   0.57   &   0.82   &   0.22  &   0.94  &   0.28  &  0.032$\pm0.010$  & IIb & IIb & Y & 0.029 & IIb & 0.029 & Y \\
1020+400  &  27.5  &   0.62   &   0.62   &   0.09  &   1.05  &   0.22  &  0.000  &   &  &  & 0.050 & IIa & & \\
1022+194  &   8.6  &   0.65   &   0.44   &   0.02  &   0.64  &   0.04  &  0.018$\pm0.007$  & I & I &  & 0.014 & I & & \\
1036-154  & -22.4  &   0.69   &   0.44   &   0.03  &   0.45  &   0.16  &  0.000  &   &  & & & & & \\
1038+528  &  40.4  &   0.61   &   0.34   &   0.05  &   0.71  &   0.20  &  0.000  &   &  & & & & & \\
1055+018  &  -4.6  &   0.69   &   2.74   &   0.19  &   3.30  &   0.35  &  0.26$\pm0.04$  & I & I & & 0.00001 & I & & \\
1100+772  &  60.7  &   0.55   &   0.41   &   0.01  &   0.19  &   0.02  &  0.000  &   &  &  & & & & \\
1116+128  &   7.5  &   0.69   &   1.98   &   0.06  &   1.46  &   0.14  &  0.058$\pm0.009$  & I &  & Y & 0.050 & IIa & & \\
1123+264  &  20.6  &   0.68   &   1.10   &   0.05  &   1.05  &   0.18  &  0.000  &   &  & & & & & \\
1127-145  & -16.5  &   0.72   &   4.94   &   0.20  &   2.55  &   0.35  &  0.000  &   &  & & & & & \\
1128+385  &  31.9  &   0.66   &   0.78   &   0.03  &   0.93  &   0.14  &  0.000  &   &  & & & & & \\
1145-071  &  -8.0  &   0.72   &   0.89   &   0.02  &   0.76  &   0.12  &  0.087$\pm0.009$  & I &  & & 0.0025 & I & & \\
1150+812  &  64.7  &   0.53   &   1.30   &   0.05  &   1.42  &   0.14  &  0.000  &   &  & &  & & 0.010 & Y \\
1155+251  &  22.5  &   0.70   &   1.23   &   0.02  &   0.71  &   0.06  &  0.000  &   &  & Y & & & 0.010 & Y \\
1200-051  &  -4.8  &   0.74   &   0.41   &   0.04  &   0.50  &   0.12  &  0.052$\pm0.007$  & I &  & & 0.00009 & I & & \\
1225+368  &  35.9  &   0.70   &   1.71   &   0.04  &   0.39  &   0.02  &  0.000  &   &  & Y & & & 0.027 & Y \\
1243-072  &  -2.4  &   0.77   &   0.61   &   0.03  &   0.72  &   0.13  &  0.121$\pm0.011$  & I & I & & 0.00001 & I & & \\
1245-197  & -13.6  &   0.78   &   3.67   &   0.05  &   1.44  &   0.15  &  0.019$\pm0.007$  & I &  & & 0.00048 & I & & \\ 
1250+568  &  54.6  &   0.67   &   1.48   &   0.03  &   0.45  &   0.04  &  NA  &   &  & & & & & \\
1253-055  &   0.2  &   0.77   &   9.28   &   0.41  &  10.87  &   1.48  &  2.3$\pm0.2$  & I & I & & 0.00001 & I & & \\
1302-102  &  -3.3  &   0.78   &   0.80   &   0.12  &   0.72  &   0.14  &  0.106$\pm0.014$  & I & I & & 0.00009 & I & & \\
1308+326  &  36.3  &   0.73   &   1.36   &   0.53  &   2.90  &   1.12  &  0.000  &   &  & & & & & \\
1328+254  &  31.9  &   0.76   &   5.11   &   0.06  &   2.25  &   0.08  &  0.000  &   &  & & & & & \\
1328+307  &  36.8  &   0.75   &  11.04   &   0.12  &   5.34  &   0.22  &  0.000  &   &  & & & & & \\
1354+195  &  29.1  &   0.78   &   1.80   &   0.09  &   1.23  &   0.20  &  0.000  &   &  & & & & & \\
1404+286  &  38.4  &   0.78   &   1.88   &   0.04  &   2.17  &   0.26  &  0.000  &   &  & & & & & \\
1409+524  &  59.0  &   0.73   &  13.53   &   0.24  &   2.07  &   0.15  &  0.000  &   &  & & 0.023 & IIa & & \\
1413+135  &  25.3  &   0.80   &   0.84   &   0.02  &   1.31  &   0.40  &  0.000  &   &  & Y & & & & \\
1430-155  &  -0.7  &   0.84   &   0.67   &   0.03  &   0.72  &   0.06  &  0.20$\pm0.03$  & I & I & Y & 0.0004 & IIa & & \\
1438+385  &  50.4  &   0.78   &   0.88   &   0.02  &   0.41  &   0.05  &  0.000  &   &  & & & & & \\
1449-012  &  14.3  &   0.84   &   0.49   &   0.02  &   0.34  &   0.04  &  0.010$\pm0.003$  & I &  & & 0.050 & IIa & & \\
1455+247  &  39.4  &   0.82   &   0.61   &   0.01  &   0.26  &   0.02  &  0.000  &   &  & Y & & & & \\
1502+106  &  26.7  &   0.84   &   2.00   &   0.13  &   1.78  &   0.43  &  0.06$\pm0.03$  & IIb? & IIb? & Y & & & & \\
1511+238  &  39.8  &   0.83   &   1.25   &   0.02  &   0.55  &   0.04  &  0.000  &   &  & Y & & & & \\
1514+197  &  36.2  &   0.84   &   0.54   &   0.07  &   0.67  &   0.10  &  0.000  &   &  & Y & & & & \\
1525+314  &  48.0  &   0.83   &   0.70   &   0.03  &   0.42  &   0.03  &  NA  &   &  & & & & & \\
1538+149  &  33.4  &   0.86   &   1.27   &   0.10  &   0.93  &   0.11  &  0.000  &   &  & Y & & & & \\
1547+508  &  67.2  &   0.81   &   0.71   &   0.02  &   1.18  &   0.12  &  NA  &   &  & & & & & \\
1555+001  &  20.0  &   0.89   &   0.74   &   0.27  &   0.76  &   0.35  &  0.000  &   &  & & & & & \\
1611+343  &  54.1  &   0.87   &   2.97   &   0.26  &   2.68  &   0.62  &  0.000  &   &  & & & & & \\
1614+051  &  25.9  &   0.90   &   0.65   &   0.03  &   0.64  &   0.07  &  0.042$\pm0.016$  & IIa? & & & & & & \\
1624+416  &  61.9  &   0.87   &   1.58   &   0.08  &   1.01  &   0.09  &  0.000  &   &  & & & & & \\
1635-035  &  18.3  &   0.92   &   0.39   &   0.05  &   0.33  &   0.03  &  0.000  &   &  & & & & \\
1641+399  &  61.1  &   0.89   &   8.70   &   0.64  &  10.96  &   2.68  &  0.000  &   &  & Y & & & & \\
1655+077  &  30.2  &   0.93   &   1.08   &   0.08  &   0.97  &   0.18  &  0.000  &   &  & & & & & \\
1656+477  &  69.4  &   0.89   &   1.22   &   0.04  &   1.24  &   0.10  &  0.058$\pm0.009$  & IIb & IIb & Y? & & & & \\
1741-038  &  19.5  &   0.96   &   2.08   &   0.29  &   2.68  &   0.38  &  0.09$\pm0.03$  & IIb & IIb & Y & 0.011 & IIb & & \\
1742-289  &  -5.6  &   0.97   &   0.46   &   0.10  &   0.81  &   0.18  &  NA  &   &  & & & & & \\
1749+096  &  33.1  &   0.97   &   1.37   &   0.18  &   2.51  &   0.85  &  0.000  &   &  & & & & & \\
1749+701  &  86.3  &   0.51   &   1.07   &   0.03  &   0.94  &   0.29  &  0.000  &   &  & & & & & \\
1756+237  &  47.2  &   0.97   &   0.81   &   0.04  &   0.59  &   0.06  &  0.039$\pm0.008$  & IIb & IIb & & 0.029 & IIb & &  \\
1803+784  &  78.1  &   0.47   &   2.39   &   0.19  &   2.88  &   0.42  &  0.11$\pm0.03$  & IIb? & IIb? & & & & & \\
1807+698  &  86.7  &   0.44   &   1.50   &   0.06  &   1.90  &   0.22  &  0.000  &   &  & Y & & & & \\
1821+107  &  34.0  &   0.99   &   1.17   &   0.05  &   0.84  &   0.11  &  0.000  &   &  & & & & & \\
1823+568  &  79.9  &   0.03   &   1.19   &   0.06  &   1.38  &   0.27  &  0.000  &   &  & & & & & \\
1828+487  &  71.8  &   0.02   &   7.99   &   0.18  &   3.54  &   0.35  &  0.000  &   &  & & 0.050 & IIa & & \\
1830+285  &  51.7  &   0.00   &   0.45   &   0.05  &   0.52  &   0.11  &  0.042$\pm0.010$  & IIb & & Y & 0.0016 & IIb & & \\
1845+797  &  76.5  &   0.45   &   0.89   &   0.32  &   0.79  &   0.06  &  NA  &   &  & & & & & \\
1928+738  &  79.6  &   0.37   &   3.76   &   0.12  &   3.99  &   0.25  &  NA  &   &  & & & & & \\
1943+228  &  43.3  &   0.07   &   0.31   &   0.02  &   0.16  &   0.02  &  NA  &   &  & & & & & \\
1947+079  &  28.6  &   0.06   &   1.29   &   0.05  &   0.72  &   0.05  &  0.000  &   &  & & & & & \\
2005+403  &  58.5  &   0.11   &   3.41   &   0.11  &   3.69  &   0.57  &  0.000  &   &  & & & & & \\
2007+776  &  75.4  &   0.40   &   1.50   &   0.22  &   2.36  &   0.46  &  0.000  &   &  & Y & & & & \\
2008-068  &  13.0  &   0.07   &   2.12   &   0.08  &   0.78  &   0.06  &  NA  &   &  & & & & & \\
2032+107  &  28.6  &   0.10   &   0.82   &   0.11  &   0.54  &   0.13  &  0.000  &   &  & & & & & \\
2037+511  &  65.2  &   0.17   &   4.85   &   0.24  &   4.05  &   0.47  &  0.000  &   &  & & & & & \\
2047+098  &  26.7  &   0.10   &   0.49   &   0.05  &   0.64  &   0.15  &  0.000  &   &  & & & & & \\
2059+034  &  19.7  &   0.11   &   0.68   &   0.08  &   0.97  &   0.29  &  0.08$\pm0.02$  & IIa & & & 0.012 & IIa & & \\
2105+420  &  55.0  &   0.16   &   1.42   &   0.04  &   0.75  &   0.10  &  0.000  &   &  & Y & & & 0.0081 & Y \\
2113+293  &  42.9  &   0.14   &   0.85   &   0.14  &   0.87  &   0.19  &  0.000  &   &  & Y & & & & \\
2121+053  &  19.8  &   0.12   &   2.27   &   0.43  &   1.72  &   0.61  &  0.000  &   &  & & & & & \\
2134+004  &  14.1  &   0.13   &   7.89   &   0.24  &   7.82  &   0.89  &  0.15$\pm0.06$  & I &  & & & & & \\
2155-152  &  -2.5  &   0.13   &   2.69   &   0.27  &   2.00  &   0.21  &  0.49$\pm0.06$  & I & I & & 0.00037 & I & & \\
2200+420  &  49.6  &   0.20   &   3.29   &   0.33  &   3.73  &   2.02  &  0.000  &   &  & Y? & 0.026 & IIb & & \\
2209+081  &  18.3  &   0.16   &   0.77   &   0.02  &   0.24  &   0.03  &  0.000  &   &  & & & & & \\
2214+350  &  42.2  &   0.20   &   0.48   &   0.08  &   0.57  &   0.06  &  0.000  &   &  & & & & & \\
2234+282  &  34.2  &   0.20   &   1.58   &   0.26  &   1.59  &   0.27  &  0.000  &   &  & & & & & \\
2251+158  &  21.3  &   0.19   &  12.07   &   0.82  &  11.25  &   3.15  &  0.000  &   &  & & & & & \\
2251+244  &  29.1  &   0.20   &   1.44   &   0.04  &   0.53  &   0.06  &  0.000  &   &  & & & & &  \\
2307+107  &  15.0  &   0.20   &   0.36   &   0.02  &   0.36  &   0.08  &  0.013$\pm0.005$  & I &  & & & & & \\
2319+272  &  28.8  &   0.23   &   0.79   &   0.04  &   0.58  &   0.07  &  0.000  &   &  & Y & & & & \\
2344+092  &  10.1  &   0.22   &   1.66   &   0.05  &   1.20  &   0.15  &  0.060$\pm0.011$  & I &  & & 0.027 & IIa & & \\
2352+495  &  45.0  &   0.28   &   2.11   &   0.06  &   1.14  &   0.09  &  0.030$\pm0.013$  & IIa? & IIa? & Y & 0.047 & IIa & & \\
\hline
\end{longtable}
\end{landscape}
\clearpage
\twocolumn
%}

\begin{acknowledgements}
We thank the anonymous referees for the useful comments and the thorough discussion of the manuscript. We thank Tim Sprenger and Laura Spitler for their suggestions, which helped to clarify and highlight important points of the article. This research is based on data of the Green Bank Interferometer (GBI), which was a facility of the National Science Foundation operated by the National Radio Astronomy Observatory under contract with the US Naval Observatory and the Naval Research Laboratory during these observations. The UMRAO observations included in this analysis were obtained as part of programs funded by a series of grants from the NSF. Additional funding for the operation of UMRAO was provided by the University of Michigan.
\end{acknowledgements}

%-------------------------------------------------------------------

\bibliographystyle{aa} % style aa.bst
\bibliography{references}

\begin{appendix}

\section{Investigation of TDV as a possible calibration problem}
\label{sec:App1}

TDV is characterised by the occurrence of simultaneous events, resembling ESEs, in different sources; correlated flux density variations in different sources may suggest the existence of an instrumental or a calibration problem. 

Given the periodic recurrence of TDV, the hypothesis of an instrumental problem seems implausible. The detection of ESE-shaped variability in 0528+134 with three different telescopes, applying different calibration schemes, during the 1993.5 event seems to rule out a calibration problem as well. However, in consideration of the ambiguity of the 0528+134 variability in 1993.5 (is it TDV or SRV?), it seems useful to analyse in some detail the variability characteristics of the calibrators, to asses the possibility of a calibration problem.

All three sources used to create a hybrid calibrator show a smooth 1-year periodic oscillation correlated with solar elongation. The minima of the oscillation occur at 0.3\,y for 0237-233, at 0.78\,y for 1245-197 and at 0.76\,y for 1328+254; they are therefore strongly out-of-phase, almost in phase opposition. A combination of smooth oscillations, however, cannot cause ESE-shaped variability episodes. Through the de-trending procedure, we isolated the fast variability component in the calibrators, to compare it with the combined 2\,GHz variability curve shown in Fig. \ref{fig:allTime}. It should be noted that 1245-197 has an ecliptic latitude of -13.6\dgr, and it is affected by type I SRV; the ESE-like variations at minimum solar elongation introduce in some sources mild 1-year periodic flares of short duration. However, this does not cause the semi-annual dips at 0.0 and 0.5\,y. 

\begin{figure}
   \centering
   \includegraphics[width=0.85\columnwidth]{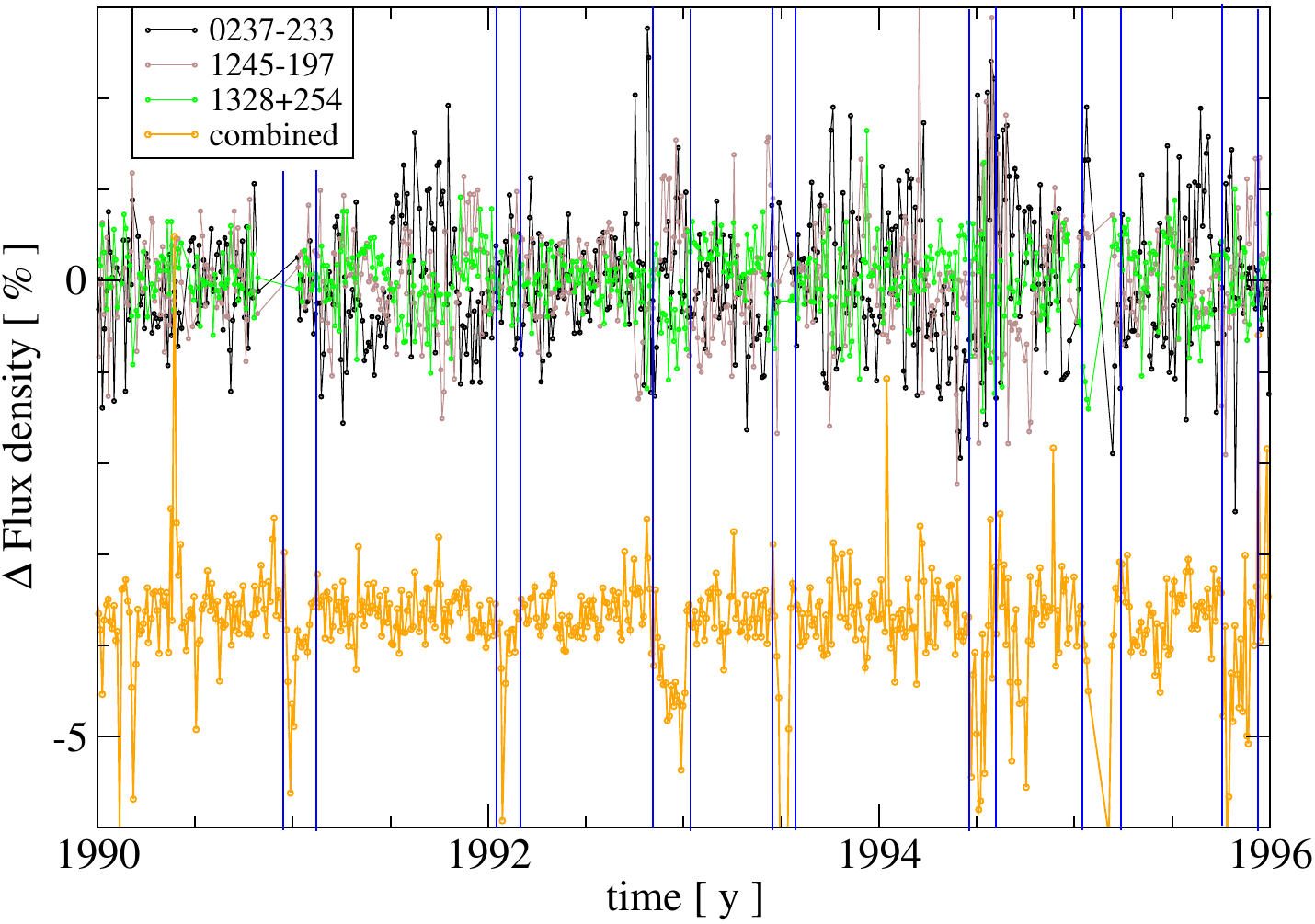}
      \caption{Comparison between the de-trended calibrators' light curves and the combined 2\,GHz variability curve plotted in Fig. \ref{fig:allTime}. Data are averaged on 0.02\,y bins; the combined variability curve is arbitrarily shifted for an easier comparison. Blue lines indicate the start and the end of the semi-annual ESE-like episodes in the combined variability curve.}
      \label{fig:cal}
\end{figure}

The 2 GHz combined variability curve and the calibrators de-trended light curves are shown in Fig. \ref{fig:cal}, after averaging them on 0.02\,y bins to reduce the random fluctuations. Given their usage in the calibration procedure, discrepancies between the calibrators' light curves tend to cancel out, which implies a significant anti-correlation between their variations; 1245-197 (brown dots) and 1328+254 (green dots) generally show a better agreement among each other than with 0237-233 (black dots), because their angular distance is much smaller, and they are observed approximately at the same time of the day. 

If TDV were caused by a calibrator, its light curve would show a systematic discrepancy with respect to the other two during the ESE-shaped events. If the effect were instead caused by different observational conditions for 0237-233 compared to the two angularly-nearby sources 1245-197 and 1328+254, we should see a periodic deviation between the former and the latter ones. In general, strong disagreements between the calibrators do not coincide with dips in the data; in Fig. \ref{fig:cal}, only two events look correlated with a significant discrepancy between the calibrators: the 1993.0 and, to a lesser extent, the 1995.1 one. In both cases, however, the agreement is good between 0237-233 and 1245-197, while the flux density of 1328+254 is considerably lower. A dip in a calibrator would introduce a peak, not a dip, in the calibrated light curves. This implies that either two distant calibrators are affected by chance by a similar problem for a few months (which seems unrealistic) or the observed drop of flux density must affect 1328+254 as well as several of the calibrated sources. In conclusion, the calibrators do not seem to cause the problem, but they might be affected by it.

\section{List of all events detected through automated analysis}
\label{sec:AppDips}

{\bf 0003+380}: 1993.77, 1993.22, 1992.22, 1991.62, 1990.62, 1990.32\\
{\bf 0003-066}: 1992.92, 1992.22, 1991.87, 1990.22, 1989.22\\
{\bf 0016+731}: 1993.22, 1992.62, 1991.77, 1991.32, 1990.47, 1989.72, 1988.92\\
{\bf 0019+058}: 1993.24, 1992.29, 1991.24\\
{\bf 0035+121}: 1993.26, 1992.26, 1991.26, 1990.26, 1989.26\\
{\bf 0035+413}: 1992.34, 1991.44, 1990.89, 1989.74\\
{\bf 0055+300}: 1993.19, 1991.19, 1990.99, 1989.74\\
{\bf 0056-001}: 1993.25, 1992.25, 1991.25, 1990.25, 1989.25\\
{\bf 0113-118}: 1991.25, 1989.25, {\bf 1988.95}\\
{\bf 0123+257}: {\bf 1988.94}\\
{\bf 0130-171}: 1989.49, {\bf 1988.94}\\
{\bf 0133+476}: 1992.33, 1990.78, 1990.48, 1990.23, 1987.73, 1986.73\\
{\bf 0134+329}: 1992.36, 1991.26, 1991.01, 1988.91\\
{\bf 0147+187}: 1993.46, 1992.31\\
{\bf 0201+113}: 1993.31, 1992.31, 1991.56, 1991.31, 1990.31, 1989.31\\
{\bf 0202+319}: {\bf 1993.48}, 1992.33, 1989.48, 1988.38, 1987.23, 1985.33, 1984.28\\
{\bf 0212+735}: 1994.15, 1993.35, {\bf 1993.00}, 1992.35, {\bf 1992.05}\\
{\bf 0224+671}: 1995.84, 1992.19, 1991.19, 1990.29, 1988.89, 1988.59, 1987.99, 1983.79, 1979.94\\
{\bf 0235+164}: 1992.34, 1991.34, 1986.69, 1985.34, 1984.09, 1982.89, 1982.04\\
{\bf 0237-233}: 1994.95, 1989.25, 1984.40, 1983.40, 1982.90\\
{\bf 0256+075}: {\bf 1993.49}, 1992.89, 1992.74, {\bf 1992.49}, 1991.84\\
{\bf 0300+470}: 1993.53, 1992.68, 1992.33, 1991.48, 1990.73, 1990.23, 1989.88, 1989.48, 1988.93, 1987.93\\ 
{\bf 0316+413}: {\bf 1991.03}, 1990.38, 1988.78\\
{\bf 0319+121}: {\bf 1993.51}, 1991.36, 1990.36, 1989.31\\
{\bf 0333+321}: 1993.74, 1993.39, 1991.84, 1987.94, 1987.44, 1986.34\\
{\bf 0336-019}: 1995.36, 1994.46, 1992.91, 1992.56, 1992.06, 1991.51, 1983.16, 1981.11\\
{\bf 0337+319}: 1993.44, {\bf 1992.94}, 1991.39, {\bf 1990.99}\\
{\bf 0355+508}: 1984.01, 1983.36, 1981.66, 1980.31, 1979.61\\
{\bf 0400+258}: {\bf 1993.50}, 1992.90, 1992.45, 1991.95, 1991.45, {\bf 1991.00}, 1990.55\\
{\bf 0403-132}: 1992.92, 1991.47\\
{\bf 0420-014}: 1992.20, 1989.25, 1987.30, 1986.65, 1985.35, 1984.40\\
{\bf 0440-003}: 1994.21, 1993.96, {\bf 1993.51}, 1993.26, 1990.46, 1989.86, 1989.41\\
{\bf 0444+634}: 1994.44, 1993.99, 1993.29, 1992.09, 1991.59, 1990.99, 1990.64, 1989.34, 1989.09\\
{\bf 0454+844}: 1993.52, 1993.27, 1990.22\\
{\bf 0500+019}: 1992.88, 1992.23\\
{\bf 0528+134}: 1993.95, {\bf 1993.50}, 1993.05, 1992.40, 1991.00, 1990.50, {\bf 1988.95}\\
{\bf 0532+826}: 1991.72, 1989.97, 1989.07\\
{\bf 0537-158}: 1993.50, 1993.35, 1993.00, 1992.45, 1991.95, 1991.45, 1990.95, 1990.50, 1989.30\\
{\bf 0538+498}: 1991.01, 1990.46\\
{\bf 0552+398}: 1993.52, 1992.17, 1991.47, 1990.62, 1988.77, 1987.47, 1986.22, 1985.82, 1984.42\\
{\bf 0555-132}: 1993.92, 1993.47, 1992.67, 1992.27, 1990.02, 1989.42\\
{\bf 0615+820}: 1993.98, 1993.53, 1992.83, 1992.08, 1991.58, 1990.98\\
{\bf 0624-058}: 1992.60, 1990.45\\
{\bf 0633+734}: {\bf 1993.48}, 1992.98, 1990.48, 1990.28\\
{\bf 0650+371}: 1992.55, 1991.65, 1991.00, 1990.45, 1989.05\\
{\bf 0653-033}: 1994.12, 1993.47, 1992.32, 1991.72, 1990.52, 1989.07, 1988.82\\
{\bf 0716+714}: 1991.40, 1990.00, 1988.75, 1988.15\\
{\bf 0723+679}: 1993.55, 1993.00, {\bf 1991.00}, 1990.45\\
{\bf 0723-008}: 1993.14, 1989.74\\
{\bf 0742+103}: 1992.95, 1992.55, 1990.55, 1986.10, 1982.20\\
{\bf 0743+259}: 1993.34, 1990.54, 1989.54\\
{\bf 0759+183}: 1993.76, 1993.51, 1992.51, 1991.86, 1990.01, 1989.56\\
{\bf 0804+499}: 1994.54, 1991.64, 1990.59\\
{\bf 0818-128}: 1992.59, 1990.04, 1989.44\\
{\bf 0827+243}: 1994.52, 1993.52, 1992.52, 1991.52\\
{\bf 0836+710}: 1995.93, 1995.38, 1994.93, 1994.53, 1993.38, 1992.98, 1991.48, 1991.03, 1990.48, 1989.58\\
{\bf 0837+035}: 1993.49, 1992.94, 1992.39\\
{\bf 0851+202}: 1991.54, 1989.59, 1986.34, 1985.14, 1983.59, 1979.54\\
{\bf 0859-140}: 1993.67, 1992.52, 1991.47, 1990.07\\
{\bf 0922+005}: 1993.58, 1993.08, 1992.48, 1991.68, 1991.13, 1990.18\\
{\bf 0923+392}: {\bf 1992.99, 1991.99}, 1991.59, 1988.69, 1987.59, 1986.04, 1984.44\\
{\bf 0938+119}: 1995.48, 1992.63\\
{\bf 0945+408}: {\bf 1993.50}, 1991.50, 1991.10\\
{\bf 0952+179}: {\bf 1993.48}, 1991.58\\
{\bf 0954+658}: 1994.52, 1992.97, 1990.47, 1989.47, 1986.67, 1981.12, 1980.37\\
{\bf 1020+400}: {\bf 1993.52}, 1990.62\\
{\bf 1022+194}: 1993.75, {\bf 1993.50}, 1992.90, 1991.65\\
{\bf 1036-154}: {\bf 1993.54}, 1992.64, 1991.14\\
{\bf 1038+528}: 1992.01\\
{\bf 1055+018}: 1991.74\\
{\bf 1100+772}: 1993.50, 1993.20, 1992.95, 1991.00\\
{\bf 1116+128}: {\bf 1993.49}, 1990.69\\
{\bf 1123+264}: {\bf 1993.53}, 1992.58, {\bf 1991.03}\\
{\bf 1127-145}: 1993.72, 1991.07\\
{\bf 1128+385}: 1990.31\\
{\bf 1145-071}: {\bf 1993.52}, 1991.62, 1989.72\\
{\bf 1150+812}: {\bf 1993.53, 1991.03}\\
{\bf 1155+251}: {\bf 1993.50, 1993.00}\\
{\bf 1200-051}: {\bf 1993.54}, 1992.04, 1991.74\\
{\bf 1225+368}: {\bf 1993.50, 1992.95}, 1991.55\\
{\bf 1243-072}: 1992.57, 1991.82, 1991.02, 1990.62, 1990.32\\
{\bf 1253-055}: 1995.62, 1993.07, 1992.77\\
{\bf 1302-102}: 1991.68, 1990.63, 1990.38\\
{\bf 1308+326}: 1993.98\\
{\bf 1328+254}: {\bf 1995.06}, 1992.96, 1991.56, 1981.11\\
{\bf 1328+307}: {\bf 1995.05}, 1985.60, 1979.75\\
{\bf 1354+195}: 1993.73, 1990.83, 1989.58\\
{\bf 1404+286}: 1994.23, {\bf 1993.53}\\
{\bf 1409+524}: 1991.78, 1990.78\\
{\bf 1413+135}: {\bf 1993.50}, 1992.85, 1991.95, 1990.85\\
{\bf 1430-155}: 1993.79, 1992.89, 1991.89, 1990.84, {\bf 1990.04}, 1989.14, 1988.84\\
{\bf 1438+385}: 1994.23, {\bf 1993.53}\\
{\bf 1449-012}: 1992.84\\
{\bf 1455+247}: {\bf 1993.52}, 1992.92\\
{\bf 1502+106}: 1994.07, 1991.24, 1990.19, 1989.79, 1987.49, 1986.19, 1985.69, 1984.84, 1980.29\\
{\bf 1511+238}: {\bf 1993.48}, 1992.93, 1991.93\\
{\bf 1514+197}: 1994.99, {\bf 1992.94}, 1992.19, 1991.44, 1990.09, 1989.44, 1989.09, 1988.79\\
{\bf 1538+149}: 1991.31, {\bf 1990.96}, 1990.16\\
{\bf 1555+001}: 1993.94, 1993.59, 1993.39, 1989.89, 1988.24, 1987.14\\
{\bf 1611+343}: 1992.97, 1991.17, 1989.92, 1987.92, 1985.67, 1985.42, 1984.42, 1983.72\\
{\bf 1614+051}: 1993.85, 1993.60, 1993.30, 1992.80, 1992.30, 1992.00, 1991.00, 1990.55, 1989.80\\
{\bf 1624+416}: 1992.92, {\bf 1989.47}\\
{\bf 1635-035}: 1993.77, 1993.22, 1992.92, 1992.67, 1991.42, 1990.67, 1989.32, 1989.02, 1988.77\\
{\bf 1641+399}: 1993.39, 1992.94, 1991.99, 1987.84, 1986.74, 1986.04, 1985.59, 1979.59\\
{\bf 1655+077}: 1993.73, 1993.43, 1992.78, 1992.18, 1991.83, 1991.53, 1989.08\\
{\bf 1656+477}: 1992.94, 1992.49, 1991.94, 1991.34\\
{\bf 1741-038}: 1993.36, 1992.91, 1992.41, 1990.21, 1989.41, 1988.41, 1987.96, 1987.36, 1986.46, 1984.91\\
{\bf 1749+096}: 1993.82, 1991.57, 1989.27, 1987.97, 1987.47, 1987.02, 1986.17, 1985.17, 1983.82\\
{\bf 1749+701}: 1993.36, 1992.96, 1989.96, 1989.51, 1988.91\\
{\bf 1756+237}: 1993.97, 1993.52, 1992.62, 1991.52, 1991.27, 1990.52, 1990.02\\
{\bf 1803+784}: 1993.87, 1992.92, 1991.72, 1990.77\\
{\bf 1807+698}: 1992.99, 1985.64\\
{\bf 1821+107}: 1993.89, 1993.54, 1991.94, 1991.39, 1988.39, 1985.29, 1984.69, 1984.24\\
{\bf 1823+568}: 1993.88, 1992.08, 1990.58\\
{\bf 1828+487}: 1991.02\\
{\bf 1830+285}: 1993.90, 1993.50, 1992.95, 1992.45, 1991.95, 1991.45, 1990.80, 1990.45, 1989.50, 1988.50\\
{\bf 1928+738}: 1993.87, 1993.62, 1993.37, 1992.97\\
{\bf 1943+228}: 1993.87, 1993.47, 1993.12\\
{\bf 1947+079}: 1993.96, 1992.71, 1991.66, 1988.91, 1988.56\\
{\bf 2007+776}: 1993.35, {\bf 1991.00}, 1990.10, 1989.45\\
{\bf 2008-068}: 1988.94\\
{\bf 2032+107}: 1993.85, 1992.95, 1992.40, 1991.90, 1991.45, {\bf 1991.00}, 1989.15\\
{\bf 2037+511}: 1991.92, 1991.52\\
{\bf 2047+098}: 1992.95, 1992.10\\
{\bf 2059+034}: {\bf 1993.51, 1992.96}, 1992.06, 1991.16, 1990.11, 1988.91\\
{\bf 2105+420}: 1992.46, 1992.06, 1991.56, {\bf 1991.01}\\
{\bf 2113+293}: 1993.09, 1992.49, 1991.54, 1989.94, 1989.14\\
{\bf 2121+053}: 1993.87, 1992.62, 1991.37, 1990.02, 1989.32\\
{\bf 2134+004}: 1993.08, 1991.18, 1987.88, 1986.63\\
{\bf 2155-152}: 1993.63, 1992.28, 1991.83, 1991.18, 1990.13\\
{\bf 2200+420}: 1992.25, 1990.55, 1988.90, 1986.70, 1985.20, 1984.50, 1981.70\\
{\bf 2209+081}: 1993.81, 1993.01, 1991.11, 1990.16\\
{\bf 2214+350}: 1992.90, 1991.45, 1989.75, 1989.45, 1988.60\\
{\bf 2234+282}: 1993.20, 1989.40, 1987.55, 1985.90, 1984.95, 1983.40\\
{\bf 2251+158}: {\bf 1992.99}, 1989.44, 1985.59, 1984.89, 1983.09, 1980.34\\
{\bf 2251+244}: {\bf 1993.50}, 1993.15, 1992.50, 1992.25, 1991.55\\
{\bf 2307+107}: 1993.70, 1989.40\\
{\bf 2319+272}: 1992.88, 1992.48, 1992.18, 1990.53, 1989.28\\
{\bf 2344+092}: 1993.42, 1992.17, 1991.22, 1990.52, 1990.17\\
{\bf 2352+495}: 1993.68, 1993.28, 1991.23, 1990.33, 1985.38, 1984.68\\

\section{Extended discussion of individual ESEs}
\label{sec:App2}

\begin{figure}
   \centering
   \includegraphics[width=0.85\columnwidth]{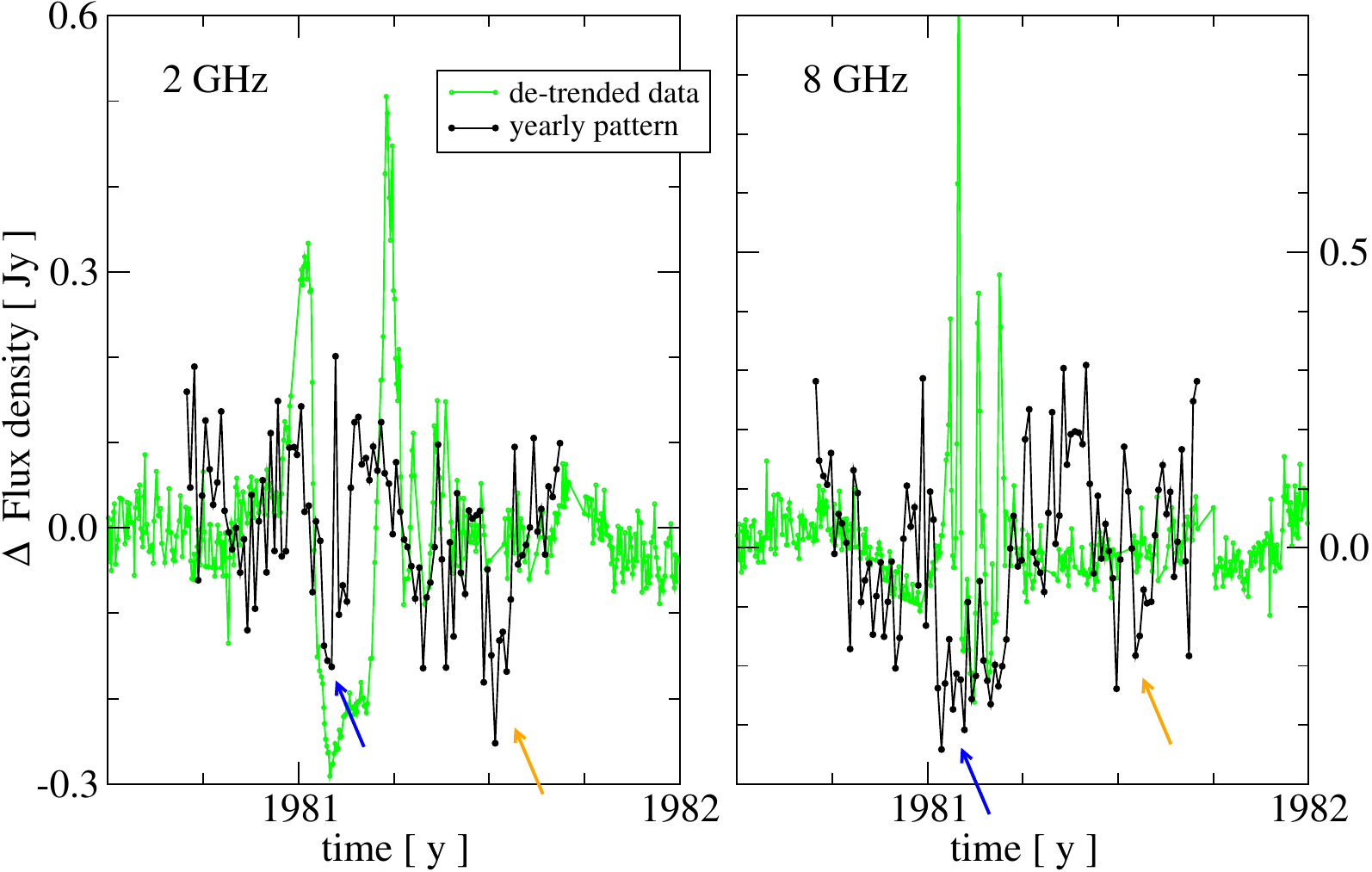}
      \caption{The de-trended 2 and 8 GHz light curves of 0954+658 (left and right panel respectively, green dots) are here plotted together with the average yearly pattern obtained after removing the data-points covering the 1981.1 event, stacking the remaining data and averaging them (black dots). The stacked data are multiplied by a factor 9 for a better comparison between the datasets. The peak-to-dip difference in the 1981.1 event is about 20 times greater than the spread in the noise level of the source. Blue and orange arrows indicate the time of maximum and minimum solar elongation of the source.
      }
      \label{fig:0954}
\end{figure}

\begin{figure}
   \centering
   \includegraphics[width=0.85\columnwidth]{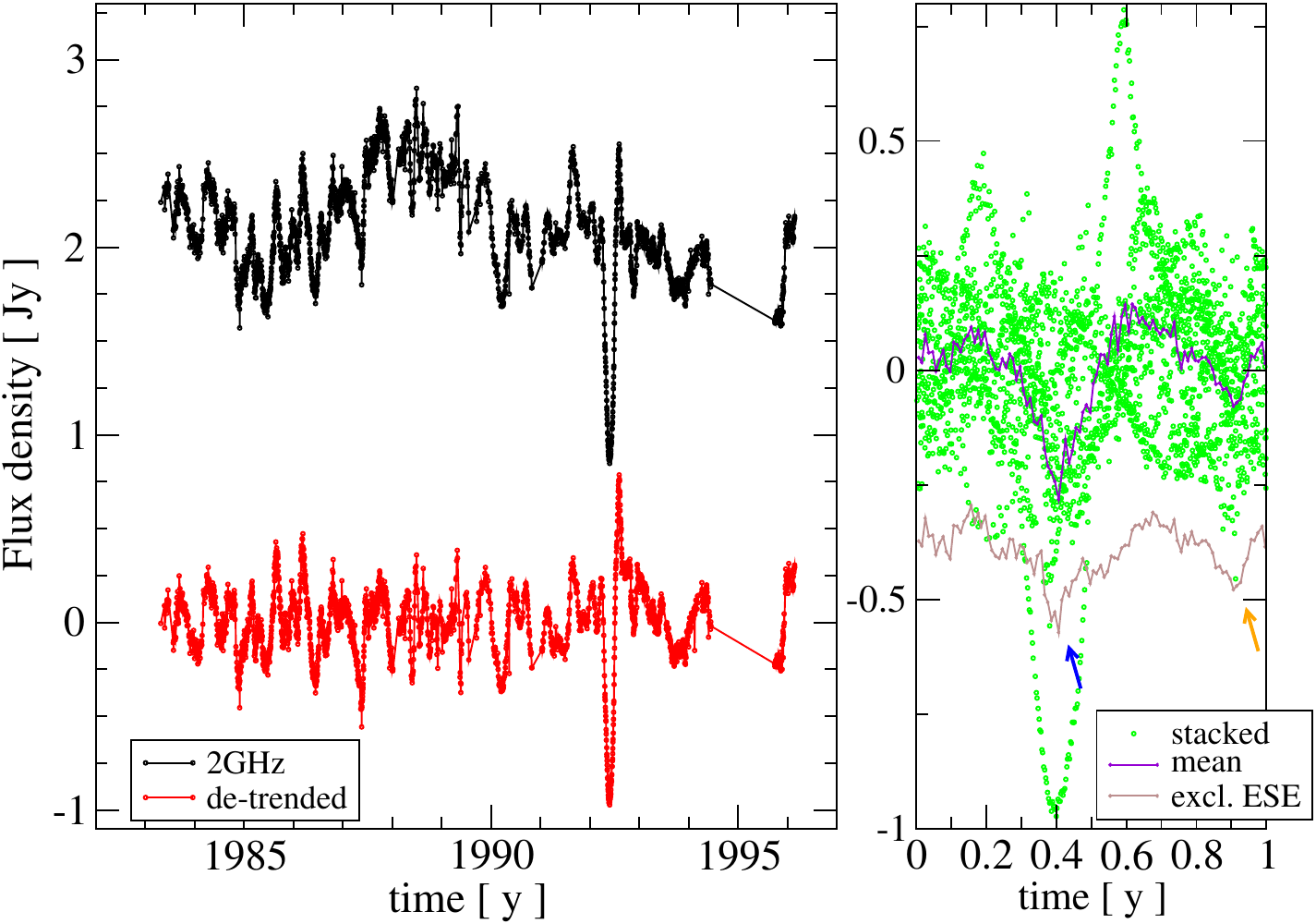}
      \caption{Left panel: in black, the 2 GHz light curve of 1741-038; the de-trended data are shown in red. Right panel: the de-trended data, stacked on one-year intervals, are shown as green dots; averaged data are shown as violet dots, while stacked data not including the 1992.5 event are shown in brown, after shifting them in flux density for an easier comparison. At 2 GHz, the peak-to-dip difference in the 1992.4 event is about 40 times greater than the spread in the noise level of the source. As in the previous figure, blue and orange arrows indicate the time of maximum and minimum solar elongation of the source.}
      \label{fig:1741}
\end{figure}

Here below we report a short analysis of the individual ESEs identified in \cite{2001ApJS..136..265L}, in the light of the SRV/TDV features discussed in the present article.
\begin{itemize}
\item The event in the light curves of 0201+113 and the two detected for 0952+179 are due to type I SRV.
\item Three events (for 0133+476, 0300+470, and 1749+096) happened during GBI transition (in 1988.1-1988.2\,y); it is plausible that their origin is instrumental.
\item The 1990.2 event in the light curves of 0133+476 (\ecl: 34.8\dgr, minimum elongation at 0.33\,y) is not easy to identify: the 2 GHz data are characterised by very fast variations whose shape resembles ESEs; much slower and less frequent variations are visible in the 8 GHz data. It is not clear whether the variations at the two frequencies are correlated at the time of the event. Since there is no clear sign of SRV, this has been classified as an ESE.
\item 0202+319 (\ecl: 18.3\dgr, minimum elongation at 0.33\,y) shows ESE-like variability in 1989.5; other episodes of similar variations can be seen at 1990.5, 1993.0,  and 1993.5, suggesting TDV as the most likely explanation.
\item The 1988.3 event in 0300+470 (\ecl: 28.7\dgr, minimum elongation at 0.38\,y) at 2 GHz has a complex shape that cannot be immediately associated with an ESE, while at 8 GHz a dip is hardly detectable. Assuming it can be classified as an event, it is unlikely to be due to SRV, as it ends before the minimum solar elongation is reached. 
\item The 1986.37 event of 0333+321 (\ecl: 12.6\dgr, minimum elongation at 0.39\,y) is certainly caused by the Sun. The source is affected by type II SRV, with clear dips, both at 2 and 8 GHz, in 1984.4, 1992.4, and 1993.4. A second ESE has been identified at 1987.95; since there is no evidence that the source is affected by TDV, we classified it as an ESE.
\item The three events in 0528+134 have been extensively discussed in Sect. \ref{sec:0537}; they are due to type IIb SRV.
\item The case of 0954+658 (\ecl: 48.8\dgr, minimum elongation at 0.57\,y) needs to be carefully considered, as its 1981.1 event represents the archetype of all ESEs. The source is affected by type IIb SRV. After removing the data covering the 1981.1 event, the data at both frequencies have been de-trended (with 0.3\,y interpolation timescale) and stacked, looking for yearly patterns; at 2GHz, the stacked data (black dots in Fig. \ref{fig:0954}, left panel; the flux has been multiplied by a factor 9 for a better comparison with the event) show a semi-annual pattern, which at the beginning of the year (blue arrow) has a clear ESE-like shape. The 1981.1 event (emphasised in the de-trended data, green dots) is much stronger than the semi-annual event in the stacked data, and it lasts longer, but they are both centred at the time of maximum solar elongation. Even more suggestive is the behaviour of the 8 GHz data (Fig. \ref{fig:0954}, right panel): here the ESE-like pattern at the beginning of the year is stronger, and has the same duration as the 1981.1 event. This strongly favours an interpretation of the event in terms of SRV.
\item The $\sim$1993.5 event in 1438+385 (\ecl: 50.4\dgr, minimum elongation at 0.78\,y) is due to TDV.
\item The yearly pattern of 1502+106 (\ecl: 26.7\dgr, minimum elongation at 0.84\,y) has a complex shape that suggests the possible influence of both type IIb SRV and TDV. The 1979.5 event is likely due to TDV; the 1986 event seems more like a sequence of two or more variability episodes, it is therefore hard to classify.
\item In the light curves of 1611+343 (\ecl: 54.1\dgr, minimum elongation at 0.87\,y) one event was detected in 1985.4, approximately at the time of maximum solar elongation. The source, however, shows one single episode of SRV in 1993.9, and it does not appear to be strongly influenced by the Sun. It is likely that the ESE is real.
\item The impressive event in 1992.5\footnote{As per definition by \citealt{2001ApJS..136..265L}; the event is centred, more precisely, on 1992.42.} in the light curves of 1741-038 (\ecl: 19.5\dgr, minimum elongation at 0.96\,y) is due to SRV. This conclusion is supported by the determination of the average yearly pattern of the variability: the original light curve (see Fig. \ref{fig:1741}, left panel, black dots) has been de-trended (red dots) and then stacked (left panel, green dots). The average of the stacked data has been calculated both including all data (violet dots) and excluding the data covering the 1992.5 event (brown dots; the data have been shifted in flux for more clarity). In both cases, two clear minima are visible, and occur very close (i.e. at 1985.42 and 1985.92) to the time of minimum and maximum solar elongation, in agreement with the characteristics of type IIb SRV.
\item  The 1993.7 event of 1756+237 (\ecl: 47.2\dgr, minimum elongation at 0.97\,y) is hard to classify. The source shows clear type IIb SRV, with regular dips around 0.0 and 0.5\,y, which though are not compatible with the time of the discussed event. On the other hand, the yearly pattern in the 2 GHz light curve of 1756+237 is more complex than for other sources, showing also regular dips at 0.7\,y. The event has been conservatively classified as an ESE.
\item The event detected for 1821+107 (\ecl: 34.0\dgr, minimum elongation at 0.99\,y) in 1984.25 cannot be explained in terms of SRV. It is worth noting that it may be part of a yearly sequence, as three additional, less prominent, dips can be found in 1985.28, 1986.24, 1987.28. However, given the high variability on short timescales, this may also happen by chance.
\item For 2251+244 (\ecl: 29.1\dgr, minimum elongation at 0.20\,y) an event is reported at 1988.9\footnote{In Table 5 of \citealt{2001ApJS..136..265L} it is reported at 1989.9, but it actually occurs in 1988.9, as shown by Fig. 7 in the paper.}. The simultaneity with similar flux density drops in other sources demonstrates that it is due to TDV. 
\item 2352+495 (\ecl: 45.0\dgr, minimum elongation at 0.28\,y) is affected by type IIa SRV, with dips around the time of minimum solar elongation in 1990.3, 1991.3, 1993.3 and 1994.3; \cite{1994ApJ...430..581F} reports an event with rather complex shape starting in 1984.5 and ending in 1985.5, while \cite{2001ApJS..136..265L} indicates as an ESE the interval between 1984.5 and 1984.8. The flux density variation could be accounted for as a sequence of two events centred in 1984.78 and 1985.3, therefore at maximum and minimum solar elongation, but no final conclusion can be drawn.
\end{itemize}

\end{appendix}

\end{document}